\documentclass[a4paper]{iopart}

\usepackage{graphicx}
\usepackage{epstopdf} 
\usepackage{amsthm} 
\newcommand{\eins}{\mbox{$1 \hspace{-1.0mm} {\bf l}$}}                 
 
\usepackage[
    bookmarks=true,         
    unicode=false,          
    pdftoolbar=true,        
    pdfmenubar=true,        
    pdffitwindow=false,     
    pdfstartview={FitH},    
    pdftitle={My title},    
    pdfauthor={Author},     
    pdfsubject={Subject},   
    pdfcreator={Creator},   
    pdfproducer={Producer}, 
    pdfkeywords={keyword1} {key2} {key3}, 
    pdfnewwindow=true,      
    colorlinks=false,       
    linkcolor=red,          
    citecolor=green,        
    filecolor=magenta,      
    urlcolor=cyan           
]{hyperref}

\newtheorem{theo}{Theorem}

\begin{document}

\topical[SSE]{A stochastic approach to open quantum systems\footnote{Published in Topical Reviews of the {\it Journal of Physics: Condensed Matter} {\bf 24}, 273201 (2012)}} 
\author{R. Biele$^{1,2}$ and R. D'Agosta$^{2,3}$}
\address{$^1$Institut f\"ur Theoretische Physik, Technische Universit\"at
Dresden, D-01062 Dresden, Germany}
\address{$^2$ ETSF Scientific Development Center,
Departamento de F\'isica de Materiales, Universidad del Pa\'is Vasco, E-20018
San Sebasti\'an, Spain}
\address{$^3$ IKERBASQUE, Basque Foundation for Science, E-48011, Bilbao, Spain}
\eads{treborB@gmx.de, roberto.dagosta@ehu.es}
\begin{abstract}
Stochastic methods are ubiquitous to a variety of fields, ranging from Physics to Economy and
Mathematics. In many cases, in the investigation of natural processes, stochasticity arises
every time one considers the dynamics of a system in contact with a somehow bigger system, an
environment, that is considered in thermal equilibrium. Any small fluctuation of the
environment has some random effect on the system. In Physics, stochastic methods have been
applied to the investigation of phase transitions, thermal and electrical noise, thermal
relaxation, quantum information, Brownian motion etc.

In this review, we will focus on the so-called stochastic Schr\"odinger equation. This is
useful as a starting point to investigate the dynamics of open quantum systems capable of
exchanging energy and momentum with an external environment. We discuss in some details the
general derivation of a stochastic Schr\"odinger equation and some of its recent applications
to spin thermal transport, thermal relaxation, and Bose-Einstein condensation. We thoroughly
discuss the advantages of this formalism with respect to the more common approach in terms of
the reduced density matrix. The applications discussed here constitute only a few examples of
a much wider range of applicability.

\end{abstract} 
\tableofcontents
\pacs{05.30.-d, 05.70.Ln, 31.10+z, 31.15.ee, 75.10.Pq, 67.85.-d} 
\maketitle


\section{Introduction}\label{introduction}
In the scientific investigation of a natural system, usually the first step of the modelling
process is to consider the system as closed and isolated. The system then evolves following
its natural laws. The direct observation of the system --our experience-- helps us in
identifying the natural laws that many systems obey. An abstraction process allows us to
formulate a coherent set of principles we believe are valid for a wide class of physical
systems, we therefore create a theory \cite{Vignale2011}. However, from the same experience we
know that no system is isolated or closed. For as small as we can think of, there is always
some leaking or coupling, let alone the observation process, that does not allow for a
complete decoupling of the dynamics of the system from the external environment. One of the
major challenges of theoretical physical modelling has been to investigate this coupling and
its effects on the ``natural'' system. This situation is even more striking for a quantum
mechanical system. Here, the interaction of the quantum system with an external device can
have dramatic effects that are neither adiabatic nor small. A text-book example is the
measurement process: we might think of the coupling between the quantum mechanical system and
the measurement tool as small as possible, but the effects on the dynamics can be as large as
changing completely the state of the quantum system. This issue is so fundamental that is in
fact one of the distinguishing factors between a classical and quantum mechanical theory.

If this is the situation, the investigation of real quantum systems might appear hopeless: the
coupling with an external environment will destroy any information we have on the system
itself replacing it with some kind of forced dynamics. However, we have to realize that in
many occasions the coupling between the system and --for example-- a thermal bath has some
random nature. From this point of view, it appears natural in order to describe the dynamics
of our quantum mechanical system coupled to an external environment, to use a stochastic
approach that will give some ``average behaviour.'' An example from classical mechanics is the
Brownian motion where a large particle in a liquid is hit multiple times by the smaller
particles forming the liquid. For this reason, the large particle appears to be ``suspended''
in the liquid. At a first observation it seems to be stationary, but a closer look reveals
that it is in constant random motion. The investigation of this motion brought Langevin to
search for an equation of motion for the large particles following the result obtained by
Einstein \cite{Einstein1905}: Langevin modelled the force from the liquid particles as a small
stochastic force with zero mean and with certain correlation properties, related to the
temperature of the liquid \cite{Langevin1908}. The success of the theory essentially opened
the field of stochastic equations. Another interesting aspect of the Langevin work must be
pointed out. Normally, the force that determines the dynamics in the Newton equation depends
solely on the position of all the particles forming the system. If this is the case, we
usually say that the system is ``closed'' and the dynamics is solely determined by the initial
conditions. There is a general consensus that, enlarging the system under investigation, one
in principle could ``close'' the equation of motion and obtain a force that is only a function
of the position of all the particles in the system and the environment. It is clear, however,
that in many cases, this is not what we want to do. On the one hand, the number of degrees of
freedom is so large to make any attempt at solving the problem beyond hope, while on the other
hand we deal with a large amount of information that is essentially useless for describing the
dynamics we are interested in. Again we can revert to the classical Brownian motion for a
simple example: to close the equation of motion we need to include the dynamics of all the
particles in the liquid, surely a large number of extra degrees of freedom. Langevin was able
to ``fold'' this extra degrees of freedom into the stochastic term in his effective equation
of motion. Further development have established a few ``standard'' ways to perform this
folding and nonetheless try to maintain the important features of the liquid dynamics.

In describing the dynamics of a quantum mechanical system coupled to an external environment
we will follow a similar approach. Our starting point will be the Schr\"odinger equation for
the system and the environment. We will then operate a selection of the relevant degrees of
freedom --obviously a utterly arbitrary step especially if the system and the environment are
made of the same interacting particles-- and integrate out those that are deemed irrelevant.
In this way, we obtain an effective Schr\"odinger equation, that will be stochastic in nature,
for the ``state'' of the system. The theory we will be developing in this way is similar to
the formalism for the density matrix. There one usually, starting from the von Neumann
equation for the total density matrix, derives an equation of motion for a reduced density
matrix, a master equation, that entails the physical information of the subsystem of interest.
We will show how it is possible to derive such an equation of motion for the reduced density
matrix from the stochastic Schr\"odinger equation. For this reason, the latter has been seen as the unraveling of a master equation.

In this review we will concentrate on stochastic Schr\"odinger equations that simulate the
average behavior of a variety of condensed matter systems interacting with their environments.
In doing so, we will build an ensemble of states and to obtain any physical quantity we will
average over this ensemble. For this reason, we do not require that a single ``state'' of the
system describes a physical quantum trajectory: for example, the state might be not normalized
at each instance of time, and moreover the time evolution might change the normalization. On
the other hand, in the quantum theory of measurement, the case where the bath is under
continuous observation with some type of measurement device has been investigated
\cite{Gardiner2000,Carmichael1993}. This analysis leads to stochastic Schr\"odinger equations
whose solutions are single system trajectories, so-called ``true'' quantum trajectories
\cite{Carmichael1993}. The theory is of great importance for designing feedback control on
open quantum systems \cite{Wiseman1994}, to exploit the localization property to reduce the
number of basis states needed to represent the state vector \cite{Rigo1997}, or to monitor the
state of a Bose-Einstein condensate \cite{Dalvit2002,Szigeti2010}. However, whether the
individual paths of the stochastic Schr\"odinger equation are true or not does not affect the
validity of the average results we are interested in. The interested reader could possibly
start from \cite{Carmichael1993} to explore the development of the continuous monitoring
theory.

One of the questions we will try to address is what are the physical conditions for the
establishment of a steady state, and if this steady state corresponds to any know thermal
equilibrium between the system and the environment. We will focus on the case in which there
is no particle exchange between the system and the environment, the latter then representing a
thermal bath able to supply energy and momentum to the system. For this reason we will talk
about thermal relaxation of the system towards some equilibrium. There are in the literature a
few examples of equations of motion for the state of the system that are build to describe the
relaxation of the system towards a steady state \cite{Razavy2006,Gisin1981}. While some of
them have been widely used to investigate the relaxation towards a steady state, they are
usually not derived but rather assumed due to some sought characteristics of the dynamics they
impose. We refer the interested reader to the available literature
\cite{Razavy2006,Gardiner2000,Weiss2007} for a more complete review of those results.

This review aims at becoming a seed for a growing research field. It is organized as follows:
in section \ref{open_quantum_systems} we will derive the stochastic Schr\"odinger and the
master equation both in the Markovian and non-Markovian approximation. We will also discuss
some of the issues in solving numerically the stochastic equation. This ingredient is
fundamental if we want to discuss some applications to real systems. Section
\ref{stochastic_dft} will present some recent results on how to simplify the numerical
investigation of many-body open quantum systems with techniques from the Density Functional
Theory \cite{Marques2006}. We will also discuss the possibility of closing the Kohn-Sham
system as suggested recently \cite{Yuen-Zhou2009a}.
                                                
Section \ref{applications} deals with some examples of application of the stochastic equation
to real systems: we will discuss the case of spin thermal transport, Bose-Einstein
condensation, thermal relaxation and the effect of electron energy dissipation on the ionic
motion.


\section{Open Quantum systems}\label{open_quantum_systems} 

The theory of open quantum systems has a long history that dates back to the beginning of
quantum theory. The inclusion of the coupling between a quantum system and an external
environment reached a certain degree of maturity with the pioneering works of Vernon and
Feynman \cite{Feynman1963} together with Caldeira and Legget \cite{Caldeira1983}. The general
idea is to derive an effective dynamics for the quantum system that takes into account the
coupling with the environment without solving the equation of motion for the environment. The
starting point has been the von Neumann equation for the density matrix, since the
latter can be easily connected to the thermodynamical properties of the system
\cite{Huang2001,Feynman}. Within the so-called master-equation formalism, an impressive number
of results have been obtained, setting it to be the ``de facto'' standard for the theory of
open systems.

More or less in parallel as what happened in the classical theory with the Langevin and the
Fokker-Plank equations, one can derive an effective dynamics for the ``state'' of the quantum
subsystem, the so-called stochastic Schr\"odinger equation (SSE). The theory began with an
attempt to, mimicking the situation of the Langevin theory of Brownian motion, introduce a
stochastic term to describe the relaxation dynamics of an open quantum system
\cite{Gisin1981, Strunz1996, Strunz1996a, Strunz1997, Diosi1997, Strunz2000, Strunz2004, DeVega2005b}. It then found application into the theory of quantum optics where the
environment is the electromagnetic radiation \cite{Plenio1998}. The stochastic equation was
meant to reproduce the dynamics derived from the density matrix after some average was taken.
Recently, Gaspard and Nagaoka \cite{Gaspard1999} formalized the theory starting with the
equation of motion for the environment, the system and their coupling. With some
approximations, which are similar to those invoked for the derivation of the master
equation, they arrived at a general expression for the stochastic Schr\"odinger equation
\cite{Gaspard1999}. The idea behind this approach is similar to the Gibbs' ensemble theory
\cite{Huang2001,Feynman,Landau5}: we build many replicas of the same system, each identified
by a certain realization of the dynamics of the environment --assumed in thermal equilibrium--
and let them evolve independently. To obtain physical quantities from this amount of
information we perform averages over the many realizations of the micro-state of the
environment.

\subsection{Stochastic Schr\"odinger Equation}\label{sse}
\begin{figure}[ht!]
	\includegraphics[width=8cm]{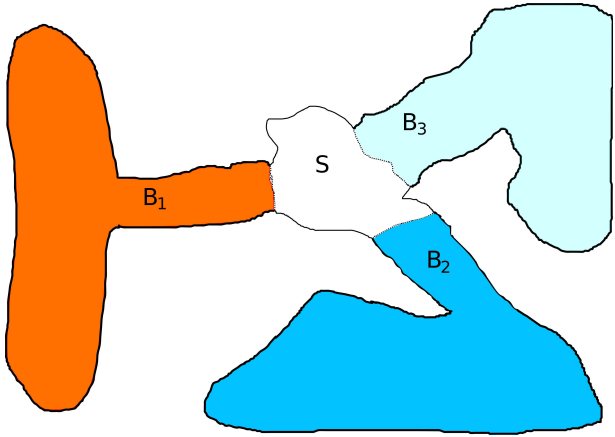}
	\caption{The subsystem $S$ is in contact with three external bath $B_1$, $B_2$, and $B_3$, that together form its ``environment''. Each of the bath is characterized by its micro-states, and possibly by a global thermal equilibrium at given temperature. We allow for energy and momentum exchange within the system and the environment.}
	\label{bathsandsystem}
\end{figure}

We begin by considering a subsystem described by the Hamiltonian $\hat{H}_S$ coupled to an
external environment, given by $\hat{H}_B$, through an interaction potential
$\lambda \hat{W}$, the Schr\"{o}dinger equation for the closed system reads
($\hbar=1$ hereafter in this review)
\begin{equation}
  \label{schroedinger_closed}
  i d |\Psi(t)\rangle= ( \hat{H}_S + \hat{H}_B +\lambda \hat{W} )
  |\Psi(t)\rangle dt .
	\end{equation}
We depict a possible situation in figure~\ref{bathsandsystem},
where a system is coupled to three different baths. Each of the baths is
characterized by its own thermodynamical micro-states as we will discuss in the
following.
	
This differential equation describes the exact dynamics of the closed system.
However, the exact microscopic description of the dynamics of the macroscopic
environment and its influence on the subsystem are in most cases not feasible.
Consequently, this equation will serve as a starting point for the derivation
of an equation of motion for the reduced wave function expressed in the
Hilbert space of the subsystem. The following deduction is very much in the
spirit of Gaspard and Nagaoka \cite{Gaspard1999}, who applied a so-called
Feshbach projection-operator method \cite{Zwanzig1960,Nakajima1958} to the
Schr\"{o}dinger equation (\ref{schroedinger_closed}) and derived a
non-Markovian stochastic Schr\"odinger equation.

By considering a complete and orthonormal basis for the environment,
\begin{equation}\label{eq:bath}
	\eins_B = \sum_{n} |n\rangle\langle n | ,~\hat{H}_B |n\rangle= \epsilon_n
|n\rangle,~\langle m|n \rangle=\delta_{mn} ,
\end{equation}
the total wave function can be expanded in this basis as
$|\Psi(t)\rangle=\sum_n |\phi^n(t)\rangle\otimes|n\rangle$. Due to the
normalisation of the total wave function,
\begin{equation}
	1=\langle\Psi|\Psi\rangle= \sum_n \langle\phi^n|\phi^n\rangle,
\end{equation}
the coefficient wave functions $|\phi^n\rangle$ are not normalised and the
square of their norm can be interpreted as the probability that the
environment is in the state $|n\rangle$. As a consequence, the wave functions
$|\phi^n\rangle$ form a statistical ensemble which determines the total state
of the combined system. In order to extract a typical representative of the
ensemble, one defines the projection operators
\begin{equation} \label{eq:PQ}
    \hat{P}\equiv\eins_S \otimes | l \rangle\langle l |,~\hat{Q}\equiv\eins_S
    \otimes \sum_{n(\neq l)} | n \rangle\langle n | .
\end{equation}
Hence, by applying the operator $\hat{P}$ to the total wave function, we extract
the $l$-th coefficient wave function, $\hat{P} |\Psi \rangle= | \phi^l\rangle
\otimes | l \rangle$. Correspondingly, $\hat{Q}|\Psi \rangle$ contains the
information of the other wave functions belonging to the ensemble. We want to
point out that these projection operators satisfy
\begin{equation}\label{eq:PQgernal}
	\hat{P}^2=\hat{P}=\hat{P}^{\dagger}, \hat{Q}^2=\hat{Q}=\hat{Q}^{\dagger}, \hat{P}+\hat{Q}= \eins_T,
\end{equation}
and thus can be used for the Feshbach projection-operator method.
Conveniently, this method is performed in the interaction picture
\begin{equation}\label{eq:schroedinger_inter}
	i d |\Psi_I(t)\rangle=  \lambda \hat{W}(t)  |\Psi_I(t)\rangle dt ,
 \end{equation}
where the total wave function and the potential in this picture are given by
\begin{eqnarray}
&& |\Psi_I(t)\rangle= e^{i \hat{H}_B t}e^{i \hat{H}_S t} |\Psi(t)\rangle,\nonumber\\ 
&& \hat{W}(t) = e^{i \hat{H}_B t}e^{i \hat{H}_S t} \hat{W} e^{-i \hat{H}_S
t}e^{-i \hat{H}_B t}.
\end{eqnarray}
The idea of the Feshbach projection method is to split the Schr\"odinger
equation for the closed system into two equations. One contains the information
about the time evolution of a typical representative of the ensemble,
$\hat{P}|\Psi \rangle$, the other is a differential equation for
$\hat{Q}|\Psi \rangle$ and describes the time evolution of all the other
coefficient wave functions. Thus, by solving the second equation and inserting
its solution into the first, one obtains a closed differential equation for
$\hat{P}|\Psi \rangle$. To follow this plan, we apply the projection operators
(\ref{eq:PQ}) to the time-dependent Schr\"odinger equation
(\ref{eq:schroedinger_inter}) and this leads to
\begin{eqnarray}
\label{eq:sub_1} 
i d \hat{P} |\Psi_I(t)\rangle& = & \lambda \hat{P}
\hat{W}(t) \hat{P} |\Psi_I(t)\rangle dt\nonumber\\ 
&&+ \lambda \hat{P} \hat{W}(t) \hat{Q} |\Psi_I(t)\rangle dt,\\
\label{eq:sub_2}
  	i d \hat{Q} |\Psi_I(t)\rangle& = & \lambda \hat{Q} \hat{W}(t) \hat{Q}
|\Psi_I(t)\rangle dt\nonumber\\ 
&&+  \lambda \hat{Q} \hat{W}(t)  \hat{P} |\Psi_I(t)\rangle dt  .
\end{eqnarray}
The second expression is an inhomogeneous linear differential equation for
$\hat{Q} |\Psi_I(t)\rangle$ and can be solved by the method of variation of
constants
\begin{eqnarray}\label{eq:sol_sub_2}
	\hat{Q} |\Psi_I(t)\rangle&=&\hat{U}(t) \hat{Q} |\Psi_I(0)\rangle\nonumber\\
	&&-i \lambda \int_0^t d t^{\prime} \hat{U}(t-t^{\prime})\hat{Q}
\hat{W}(t^{\prime})\hat{P}|\Psi_I(t^{\prime})\rangle,
\end{eqnarray}
where $\hat{U}(t)$ is the time-evolution operator of the corresponding
homogeneous differential equation and thus obeys $i d \hat{U}(t) = \lambda
\hat{Q} \hat{W}(t) \hat{Q} \hat{U}(t)dt$. Inserting (\ref{eq:sol_sub_2})
into (\ref{eq:sub_1}) leads to a closed differential equation for
$\hat{P}|\Psi_I\rangle$,
\begin{eqnarray}\label{eq:sub_1_2}
	i d  \hat{P} |\Psi_I(t)\rangle& = & \lambda \hat P \hat W (t)   \hat P|\Psi_I(t)\rangle dt  \nonumber\\
	&&+\lambda \hat P \hat W(t)   \bigg( \hat U(t)  \hat Q |\Psi_I(0)\rangle\nonumber \\ 
	&& -i \lambda \int_0 ^t d t^{\prime} \hat U(t-t^{\prime})\hat Q \hat W (t^{\prime})   \hat P |\Psi_I(t^{\prime})\rangle\bigg)dt. 
\end{eqnarray}
It is worth mentioning that until now no approximations have been made, so
that (\ref{eq:sub_1_2}) describes the exact time evolution of the $l$-th
coefficient of the total wave function. Hence, (\ref{eq:sub_1_2}) can also
be considered as a suitable starting point for a derivation of a stochastic
Schr\"{o}dinger equation beyond the weak coupling approximation. Unfortunately , this
equation is as difficult to solve as the Schr\"{o}dinger
equation for the closed system (\ref{schroedinger_closed}) and thus we will
consider a subsystem which is weakly coupled to the environment. To this end,
we perform a perturbation expansion to second order in the coupling parameter
$\lambda$,
\begin{eqnarray}\label{eq:sub_1_pert}
	i d\hat P |\Psi_I(t)\rangle= &  \lambda \hat P \hat W (t)   \hat P |\Psi_I(t)\rangle dt + \lambda \hat P \hat W (t)    \hat Q |\Psi_I(0)\rangle dt\nonumber\\
	& -i \lambda^2 \hat P \hat  W(t)  \int_0 ^t d t^{\prime} \Bigg(  \hat Q \hat W(t^{\prime})  \hat Q|\Psi_I(0)\rangle\nonumber \\
	& +  \hat Q \hat W (t^{\prime})  \hat P|\Psi_I(t^{\prime})\rangle\Bigg)dt + \mathcal{O}(\lambda^3) ,
\end{eqnarray}
where the time-evolution operator $\hat{U}(t)$ has been expanded to second
order in $\lambda$. Until now the derivation was quite generic, no restriction
on the form of the interaction potential $\hat W$ or the Hamiltonians of the
subsystem and environment has been imposed. In the following we assume the
interaction potential to be of linear form, $\hat W = \sum_{\alpha} \hat
V_{\alpha} \hat B_{\alpha}$, in the operators $\hat V_{\alpha}$ and $\hat
B_{\alpha}$ of the subsystem and the environment, respectively. These operators
can always be redefined as Hermitian operators \cite{Gaspard1999},
thus we use $\hat V^{\dagger}_{\alpha}=\hat V_{\alpha}$
and $\hat B^{\dagger}_{\alpha}=\hat B_{\alpha}$ in the following. If needed,
this restriction can easily be lifted.

By multiplying (\ref{eq:sub_1_pert}) from the left with $\langle l |$ and
assuming that $\langle l |\hat B_{\alpha}(t)| l \rangle$
vanishes \footnote{This condition can be either fulfilled through a
redefinition of the systems Hamiltonian $\hat{H}_S$ or through the choice of
the operators $\hat B_{\alpha}$.}, it simplifies to
\begin{eqnarray}\label{eq:sub_1_pert_basis}
	i d |\phi^l_I(t)\rangle&=&  |f(t)\rangle dt\nonumber\\
	&&-i \lambda^2 \sum_{\alpha , \beta}\hat V_{\alpha}(t)  \int_0^t d t^{\prime}  \hat V_{\beta}(t^{\prime})\\
	&&\times\langle l | \hat B_{\alpha}(t)  \hat B_{\beta}(t^{\prime}) | l
	\rangle|\phi^l_I(t^{\prime})\rangle dt.\nonumber
\end{eqnarray}
The forcing term
\begin{eqnarray}\label{eq:forcing_1}
  	|f(t)\rangle &=&   \lambda \sum_{\alpha,\beta,n (\neq l)} \hat V_{\alpha}(t)\Big\{\langle l |\hat B_{\alpha}(t)| n \rangle \nonumber\\
	&&-i \lambda \int_0 ^t d t^{\prime}  \hat V_{\beta}(t^{\prime}) \langle l | \hat B_{\alpha}(t) \hat B_{\beta}(t^{\prime}) | n \rangle\Big\} |\phi^n_I(0)\rangle
\end{eqnarray}
describes the influence of all the other bath modes on the $l$-th coefficient
wave function and one sees that the initial conditions $|\phi^n(0)\rangle$
enter here as an essential ingredient. By assuming that at $t=0$ the subsystem is
in a pure state and the bath is in thermal equilibrium, the total density
operator can be written as
\begin{equation}\label{eq:newnew}
	\hat{\rho}_T(0)=|\phi(0)\rangle\langle\phi(0)| \otimes \hat \rho_{B}^{\mathrm{eq}}=|\phi(0)\rangle\langle\phi(0)| \otimes \frac{e^{-\beta^{\prime} \hat H_B}}{Z_B},
\end{equation}
where $\beta^{\prime}$ is the inverse of the temperature and
$Z_B=\Tr_B{e^{-\beta^{\prime} \hat H_B}}$. Nevertheless, we are interested in
the initial wave function corresponding to this density operator. This can be
established under the assumption that the initial condition for the total wave
function is given by
\begin{equation}\label{initial_wf}
	| \Psi(0)\rangle = | \phi(0)\rangle\otimes \sum_n \sqrt{\frac{e^{-\beta^{\prime} \epsilon_n}}{Z_B}} e^{i \theta_n}| n \rangle,
\end{equation}
where $\{\theta_n\}$ are independent random phases uniformly distributed over
the interval $[0,2\pi]$. Here, we want to point out the difference in the
description of an open quantum system by wave-function or master
equation methods: random phase factors have to be considered in the initial
conditions. Due to this, the initial conditions can be written as
\begin{eqnarray}
	|\phi^n(0)\rangle&=& \langle n    | \Psi(0)\rangle = | \phi(0)\rangle\sqrt{\frac{e^{-\beta^{\prime} \epsilon_n}}{Z_B}} e^{i \theta_n}\nonumber \\
	&=&| \phi^l(0)\rangle  e^{-\frac{\beta^{\prime}}{2} (\epsilon_n-\epsilon_l)} e^{i (\theta_n-\theta_l)},
\end{eqnarray}
where we have used the fact that all coefficient wave functions at $t=0$ are
proportional to the same state $|\phi(0)\rangle$ of the subsystem and thus they can
be expressed in terms of the $l$-th.
With the help of this, the forcing term (\ref{eq:forcing_1}) can be simplified further
\begin{eqnarray}
|f(t)\rangle &\approx&\lambda \sum_{\alpha, n (\neq l)} \hat V_{\alpha}(t) \Big\langle l \Big| \hat B_{\alpha}(t) \Big[|\phi^l(0)\rangle\nonumber\phantom{\int_0^t}\\
	&&-i\lambda \sum_{\beta} \int_0 ^t d t^{\prime}  \hat V_{\beta}(t^{\prime}) \hat B_{\beta}(t^{\prime}) |\phi^l(0)\rangle\Big] \Big| n \Big\rangle e^{-\frac{\beta'}{2}(\epsilon_n-\epsilon_l)} e^{i(\theta_n-\theta_l)}\nonumber \\
  	&\approx&  \lambda \sum_{\alpha, n (\neq l)} \hat V_{\alpha}(t) \langle l |\hat B_{\alpha}(t)| n \rangle |\phi^l_I(t) e^{-\frac{\beta'}{2} (\epsilon_n-\epsilon_l)} e^{i(\theta_n-\theta_l)} \nonumber \\
  	&=&  \lambda \sum_{\alpha} \gamma^l_{\alpha}(t)  \hat V_{\alpha}(t)  |\phi^l_I(t)\rangle.
\end{eqnarray}
Here, we have used that the expression in the square brackets gives the time evolution of the
$l$-th coefficient wave function in the interaction picture up to second order in $\lambda$
for (\ref{eq:sub_1_pert_basis}). Besides, we have included the stochastic noise in
\begin{equation}
	\gamma^l_{\alpha}(t) = \sum_{n(\neq l)}  \langle l |\hat B_{\alpha}(t)| n \rangle e^{-\frac{\beta^{\prime}}{2} (\epsilon_n-\epsilon_l)} e^{i(\theta_n-\theta_l)}  ,
\end{equation}
which depends on a specific coefficient wave function. In order to eliminate
this dependence, a thermal average is performed,
\begin{eqnarray}
	\gamma_{\alpha}(t)&=&\sum_l \frac{e^{-\frac{\beta^{\prime}}{2}\epsilon_l}}{\sqrt{Z_B}}  \gamma^l_{\alpha}(t) \nonumber\\
	&=& \frac{1}{\sqrt{Z_B}}
	\sum_{l, n(\neq l)} \langle l |\hat B_{\alpha}(t)| n \rangle e^{-\frac{\beta^{\prime}}{2}\epsilon_n} e^{i(\theta_n-\theta_l)}  .
\end{eqnarray}
Additionally, one assumes that the expectation value of an operator from the bath for a typical eigenstate
$|l\rangle$ is approximately equivalent to a thermal average of the temperature of the bath,
\begin{equation}\label{def:C}
	\langle l | \hat B_{\alpha}(t)  \hat B_{\beta}(t^{\prime}) | l \rangle\approx \Tr_B \Big[ \hat{\rho}^{\mathrm{eq}}_B \hat B_{\alpha}(t)  \hat B_{\beta}(t^{\prime}) \Big] \equiv
	C_{\alpha,\beta}(t-t^{\prime}) .
\end{equation}      
For a more complete analysis on the validity of this assumption see
\cite{Berry1977,Voros1976,Voros1977,Shapiro1988,Zelti1987,Hortikar1998}.
In this expression we have defined the bath correlation function $C_{\alpha,\beta}$, which describes
the influence of the bath onto the system by tracing out the bath degrees of freedom in the dynamics.

In addition, if the bath is large enough, $\gamma_{\alpha}(t)$ consists of a sum of many complex oscillating
terms which leads to random Gaussian behaviour according to the central limit theorem. Hence, the noise is
characterised by its mean value and its variance,
\begin{equation}\label{prop_noise}
  	\overline{\gamma_{\alpha}(t)}=0,~
  	\overline{\gamma_{\alpha}(t)\gamma_{\beta}(t^{\prime})} = 0,~
  	\overline{\gamma^{\ast}_{\alpha}(t)\gamma_{\beta}(t^{\prime})} =  C_{\alpha,\beta}(t-t^{\prime}),
\end{equation}
where the relations $\overline{e^{i(\theta_n+\theta_m)}}=0$, $\overline{e^{i(\theta_n-\theta_m)}}=\delta_{nm}$
and $\langle l |\hat B_{\alpha}(t)| l \rangle=0$ have been used. We want to point out that the noise
and the bath correlation function are not independent, more precisely, the covariance function of
the noise is given by the bath correlation function. Collecting all the information, transforming
back into the partial Schr\"{o}dinger picture of the system and setting $\tau=t-t^{\prime}$,
(\ref{eq:sub_1_pert_basis}) can be written as
\begin{eqnarray}\label{eq:NMSSE}
  	i d |\phi(t)\rangle&=& \hat H_S |\phi(t)\rangle dt +\lambda \sum_{\alpha} \gamma_{\alpha}(t)  \hat V_{\alpha}  |\phi(t)\rangle dt\nonumber\\
	&& - i \lambda^2 \sum_{\alpha, \beta}\hat V_{\alpha}  \int_0 ^t d \tau  e^{- i \hat H_S\tau} \hat V_{\beta}  C_{\alpha,\beta}(\tau)|\phi(t-\tau)\rangle dt.
\end{eqnarray}
Here we have suppressed the index $l$, since we assume this wave function is a ``typical representative'' of the dynamics of the system. This again
corresponds to the Gibbs ensemble theory: with probability close to 1, we are sure that picking at random one of the
coefficient wave function, it will evolve according to (\ref{eq:NMSSE}). In this equation the change of the wave function
at time $t$ depends not only on the current state $|\phi(t)\rangle$ but also on an integral over the whole history of the
state in the interval $[0,t]$. This behaviour is called non-Markovian and thus (\ref{eq:NMSSE}) is denominated non-Markovian
stochastic Schr\"{o}dinger equation (NMSSE).

In (\ref{eq:NMSSE}) the coupling of the bath to the subsystem is described in an approximate
manner and enters in the NMSSE through the bath correlation function $C_{\alpha,\beta}(t)$ and
the stochastic noises $\gamma_{\alpha}(t)$ with the properties (\ref{prop_noise}). Thus, all
the information about the time evolution of the bath and its coupling to the subsystem is
included in the bath correlation function. In most quantum optics cases the dependence on the
past of the wave function can be neglected. This is due to the fact that the bath correlation
function decays rapidly to zero on a time scale on which the system wave function does not vary significantly. For convenience, one neglects the non-Markovian behaviour by approximating the time dependence of the bath correlation function by a $\delta$-function,
$C_{\alpha,\beta}(t-t^{\prime}) \approx D_{\alpha,\beta} \delta(t-t^{\prime})$, known under
the name of a $\delta$-correlated bath approximation. As a result, the NMSSE reduces to
\begin{eqnarray}\label{eq:MSSE}
  	i d |\phi(t)\rangle&=& \bigg[ \hat H_S  +\lambda \sum_{\alpha} \gamma_{\alpha}(t)  \hat V_{\alpha} 
	- \frac{i \lambda^2}{2} \sum_{\alpha, \beta}\hat V_{\alpha} \hat V_{\beta}  D_{\alpha,\beta}   \bigg]\nonumber\\
	&&\times|\phi(t)\rangle dt .
\end{eqnarray}             
where $\gamma_{\alpha}(t)$ are white-noise processes with $\overline{\gamma_{\alpha}(t)}=0$
and $\overline{\gamma^{\ast}_{\alpha}(t)\gamma_{\beta}(t^{\prime})} =D_{\alpha,\beta}
\delta(t-t^{\prime})$. With the help of a unitary transformation $U_{\gamma,\delta}$, that
diagonalizes $D_{\alpha,\beta}$ with corresponding eigenvalues $d_{\alpha}$, (\ref{eq:MSSE})
can be written in an It\^o differential form
\begin{equation}\label{MSSE_ito}
	d  |\phi(t)\rangle= \bigg( \big[-i\hat H_S-\frac{1}{2}\sum_{\alpha}\hat S^{\dagger}_{\alpha}\hat S_{\alpha}\big]dt+\sum_{\alpha}\hat S_{\alpha}dW_{\alpha}\bigg)  |\phi(t)\rangle,
\end{equation}
where the new set of bath operators is given by
$\hat S_{\alpha}=\lambda \sqrt{d_{\alpha}} \sum_{\beta} U_{\alpha,\beta} \hat V_{\beta}$.
Furthermore, the stochastic processes are included in
$dW_{\alpha}=-i / \sqrt{d_{\alpha}}\sum_{j}U^{\dagger}_{j,\alpha}\gamma_{j}dt$
and one can show that these satisfy the properties
\begin{equation}
	\overline{dW_{\alpha}}=0,~\overline{dW_{\alpha} dW^{\ast}_{\beta}} =\delta_{\alpha,\beta} dt.
\end{equation}
We will call (\ref{MSSE_ito}) Markovian stochastic Schr\"{o}dinger
equation (MSSE) and we want to point out that this equation does not follow
standard rules of calculus. The state $|\phi(t)\rangle$ is a stochastic
function and its time derivation is not defined at any instant of time. In
addition, the differential noise $dW$ scales on average as
$\sqrt{dt}$ and thus can be interpreted as a differential increment
of an underlying Wiener process. As a result, this differential equation is
not tractable with standard calculus and the rules have to be modifies
according to the \textit{It\^o calculus}. Since this will bring us too far from
the scope of this review, here we will only state the important results of the
It\^o calculus needed in the following sections. For a more complete treatment
of the It\^o formalism one can consult the vast literature on the subject,
here we just point out a few standard references
\cite{Karatzas1991,Protter2004}.
An important result from this calculus is the It\^o chain rule \cite{Gardiner2000,Gardiner2004},
\begin{equation}  
\label{ito_product}
d \big( |\phi \rangle  \langle \psi|  \big)= \big(d	|\phi \rangle\big)\langle \psi|  +|\phi \rangle\big(d \langle \psi| \big) + \big(d	|\phi \rangle\big)\big(d\langle \psi|\big) ,
\end{equation}
where $\phi $ and $\psi$ are two states evolving according to the MSSE (\ref{MSSE_ito}). In addition, the rules for the It\^o differentials
\begin{equation}\label{noise_ito}
	\overline{dW_{\alpha} dW^{\ast}_{\beta}}  = \delta_{\alpha,\beta}dt ,~
	\overline{dW_{\alpha} dW_{\beta}}  = 0 ,~
	\overline{dW_{\alpha} dt}  = 0 ,
\end{equation}
should be kept in mind. In the following section we will apply these It\^o
rules in order to derive the master equation that corresponds to the MSSE. In
the case of the non-Markovian SSE, where one is confronted with non-white
noises, It\^o calculus cannot be applied and one has to find another approach
to derive the corresponding master equation.


\subsection{Density Matrix formalism}\label{density_matrix} 
In the previous section we have discussed the dynamics of a quantum mechanical system
coupled to an external environment from the point of view of what we have identified
as the ``state" of the system, $|\phi\rangle$. However, for historical and practical
reasons this is not the standard starting point. It has been easier, as we will discuss
in a moment, to start from the dynamics of the reduced density matrix or statistical
operator of the system, defined as
\begin{equation}
	\hat\rho\equiv\overline{|\phi\rangle\langle\phi|}.
	\label{density-matrix}
\end{equation}

To obtain the equation of motion for the density matrix that corresponds to the Markovian SSE
(\ref{MSSE_ito}), we can start from (\ref{density-matrix}) and calculate the differential
\begin{equation}
	d\hat \rho  = d \overline{|\phi\rangle\langle\phi|}  =\overline{\big(d	|\phi \rangle\big)\langle \phi|  + |\phi \rangle\big(d \langle \phi| \big) + \big(d	|\phi \rangle\big)\big(d\langle \phi|\big)}. 
\end{equation}
Unlike in normal calculus, one also has to keep the term $d |\phi \rangle d\langle \phi|$
as it contributes on average to first order in $dt$. In the Markovian case, where
one has to deal with white-noise processes, we can apply the It\^o rules (\ref{ito_product})
and (\ref{noise_ito}) which lead us to
\begin{eqnarray}\label{lindblad}
	d\hat\rho&=&
	\overline{\Bigg[ -i \hat H_S dt +\sum_{\alpha} \Big\{ -\frac{1}{2}\hat S^{\dagger}_{\alpha}\hat S_{\alpha}dt+\hat S_{\alpha}dW_{\alpha} \Big\}\bigg]\hat{\rho} } \\
	 		 &&+\overline{\hat{\rho} \Bigg[  \hat H_S dt +\sum_{\beta} \Big\{ (-\frac{1}{2}\hat S^{\dagger}_{\beta}\hat S_{\beta} dt+\hat S^{\dagger}_{\beta}dW^{\ast}_{\beta} \Big\}\bigg] }\nonumber \\
	 	&&+ \sum_{\alpha, \beta}\overline{\hat S_{\alpha}dW_{\alpha} \hat{\rho} \hat S^{\dagger}_{\beta}dW^{\ast}_{\beta}}+\mathcal{O}(dt^2)\nonumber\\
&=&-i [\hat H_S,\hat{\rho}]dt-\frac{1}{2}\sum_{\alpha}\bigg\{ \hat S^{\dagger}_{\alpha}\hat S_{\alpha}\hat{\rho} +\hat{\rho} \hat S^{\dagger}_{\alpha}\hat S_{\alpha} -2 \hat S_{\alpha}\hat{\rho} \hat S^{\dagger}_{\alpha}\bigg\}dt+\mathcal{O}(dt^2) .\nonumber
\end{eqnarray}
This well-known Lindblad master equation is the most general type of a Markovian master
equation which is known to preserve not only the norm but also positivity and hermiticity
\cite{Lindblad1976}. This means that the Markovian SSE (\ref{MSSE_ito}) describes on average
an open system dynamics which coincide with the Lindblad dynamics. Here, we have assumed that
the Hamiltonian of the system is non-stochastic. However, if the Hamiltonian is stochastic,
one has to deal with an ensemble of Hamiltonians and the corresponding equation of motion will
most likely differ from the Lindblad master equation. We will discuss this issue in depth in
section \ref{equivalence} where a gas of interacting bosons will be considered.

As mentioned before, It\^o stochastic calculus is only applicable for white-noise processes.
However, in the non-Markovian SSE (\ref{eq:NMSSE}) one encounters with coloured noise which
is not $\delta$-correlated. A stochastic calculus for coloured noise is not so extensively
investigated as the It\^o calculus. Hence, one has to find an alternative procedure of deriving
an equation of motion for the density operator that corresponds to the NMSSE. To this end, a
comparable master equation for the NMSSE up to fourth order in $\lambda$ can be derived by
performing a perturbative expansion in $\lambda$ to arrive at \cite{Gaspard1999}
\begin{eqnarray}
\label{eq:master_2n_order}
\frac{d \hat{\rho}(t)}{dt}=&-i [\hat H_S,\hat \rho(t)]\\
 &+\lambda ^2 \int_0^{t} d \tau \sum_{\alpha,\beta} C^{\ast}_{\alpha,\beta}(\tau) \hat V_{\alpha}\hat{\rho}(t)e^{- i \hat H_S\tau}
\hat V_{\beta} e^{i \hat H_S\tau} 
\nonumber\\
 &+\lambda ^2 \int_0^{t} d \tau \sum_{\alpha,\beta}C_{\alpha,\beta}(\tau) e^{- i \hat H_S\tau} \hat V_{\beta} e^{i \hat H_S\tau}
\hat \rho(t)\hat V_{\alpha}  
\nonumber\\
&-\lambda ^2 \int_0^{t} d \tau \sum_{\alpha,\beta}C_{\alpha,\beta}(\tau)\hat V_{\alpha} e^{- i \hat H_S\tau} \hat V_{\beta} e^{i
\hat H_S\tau} \hat\rho(t)  
\nonumber\\
 &-\lambda ^2 \int_0^{t} d \tau\sum_{\alpha,\beta} C^{\ast}_{\alpha,\beta}(\tau)\hat{\rho}(t) e^{- i \hat H_S\tau} \hat V_{\beta} e^{i
\hat H_S\tau} \hat V_{\alpha}
+\mathcal{O}(\lambda^4)\nonumber .
\end{eqnarray}
In the following, we will call this equation non-Markovian master equation as
it corresponds to the NMSSE. Also here, we have assumed that the Hamiltonian is non-stochastic. In the limit $t\rightarrow \infty$ in the history term, the non-Markovian master
equation simplifies to 
\begin{eqnarray}
\label{eq:REDF}
\frac{d \hat\rho (t)}{dt}=& -i [\hat H_S,\hat\rho (t)]\\
&+\sum_{\alpha}\bigg(\hat K_{\alpha}\hat{\rho}(t)\hat V_{\alpha}+\hat V_{\alpha} \hat \rho (t) \hat K^{\dagger}_{\alpha} - \hat V_{\alpha} \hat K_{\alpha} \hat \rho (t)-
\hat \rho (t) \hat K^{ \dagger}_{\alpha} \hat V_{\alpha} \bigg),\nonumber
\end{eqnarray}
where
\begin{equation}\hat{K}_{\alpha}=\lambda^2\sum_{\beta}\int_0^{\infty} d \tau C_{\alpha,\beta}(\tau)e^{- i \hat H_S\tau} \hat V_{\beta} e^{i \hat H_S\tau} .
\end{equation}
This time-local evolution is the well-known Redfield master equation \cite{Redfield1957} and
has been widely applied to systems where the dynamics of the environment is much faster than
the system dynamics. This equation in general provides information on the long time relaxation
dynamics of the density matrix. In conclusion, we have seen that the Markovian SSE describes
on average a dynamics that corresponds to the one obtained from a Lindblad master equation and
the non-Markovian SSE corresponds to the master equation (\ref{eq:master_2n_order}), if the Hamiltonian is not stochastic. Hence,
the SSEs can also be seen as an unraveling of the corresponding master equation, i.e., a quick
and dirty way to obtain the solution of a master equation, especially when the systems size
grows one expect a better scaling behaviour of the SSEs.

\subsection{Equation of motion for observables}\label{observable-equation} 
\begin{figure}[ht!]
	\includegraphics[width=10cm]{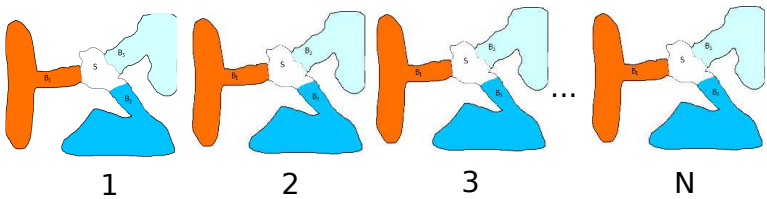}
	\caption{In evolving the SSE we build many replicas of the same system and let them follow their own time evolution. In building up the observables, we average over the contribution to the observable of each replica, as in (\ref{expectation_value}).}
	\label{replicas}
\end{figure}

Given the equation of motion for the density matrix and the SSE, either in the non-Markovian
or in the Markovian approximation, the next step is to derive a general equation of motion
for the expectation value of any observable. By definition, given a physical observable $P$,
described by the operator $\hat P$, its expectation value is given by
\begin{eqnarray}
	\label{expectation_value}
	\overline{\langle \hat P\rangle}&\equiv&\frac{\overline{\langle\phi(t)|\hat P|\phi(t)\rangle}}{{\overline{\langle\phi(t)|\phi(t)\rangle}}}=\Tr_S(\hat \rho(t)\hat P)\\
	&=&\lim_{N\to\infty}\frac{1}{N}\sum_i^N\frac{\langle\phi_i(t)|\hat P|\phi_i(t)\rangle}{\langle\phi_i(t)|\phi_i(t)\rangle},\nonumber        
\end{eqnarray}       
where, again, the $\langle \ldots\rangle$ represents the standard quantum mechanical average and $\overline{\cdots}$ the average over the noise as we have introduced them in section \ref{sse}. To understand this last process, we can think of (\ref{expectation_value}) as having build a certain number of replicas of the system, labelled by the index $i$, $\phi_i(t)$, see figure \ref{replicas}. Each replica evolves according to the SSE with a given realization of the noise. Averaging over the noise therefore corresponds to summing up all the weighted contributions to the given observable coming from each replica.    

It should be evident from (\ref{expectation_value}) why the density matrix got a relevant
amount of attention: if one were able to solve the equation of motion for the density with the
proper initial conditions, the expectation value of any physical observable would come
essentially at no extra cost. From (\ref{density-matrix}), it is clear that the process of
averaging over the many realizations of the stochastic behavior that is driving the dynamics
of $|\Psi\rangle$ is a crucial part of the definition of the density matrix. It is this
average that builds up the correlation between the different trajectories of the stochastic
dynamics. The same average step makes the SSE difficult to handle: even if one were able to
write down an explicit solution to the SSE, the physical quantities will appear only after the
average over many realizations of the noise is taken. For the density matrix, this step is
embedded in the definition, and is thus carried out before the dynamics is solved, obtaining
therefore an equation for the averaged quantities.

We want to derive an equation of motion for $\langle \hat P\rangle$. To this end, we consider
the case of the Lindblad equation or the MSSE (\ref{MSSE_ito}). The generalization to the
non-Markovian dynamics is lengthly involving the expansion of the equation of motion in terms
of the coupling parameter $\lambda$ \cite{Gaspard1999}.
                                           
Considering the case of a single Wiener process, $dW$, we have by using It\^o calculus
\begin{eqnarray}
d\langle\hat P\rangle&=&(d\langle\phi|)\hat P|\phi\rangle+\langle
\phi|\hat P(d|\phi\rangle)+ (d\langle\phi|)\hat P(d|\phi\rangle)\nonumber\\
&=&\left\langle-i[\hat P,\hat H]-\frac12 \left( \hat S^\dagger \hat S
\hat P+\hat P \hat S^\dagger
\hat S\right)\right\rangle dt+\left\langle\hat S^\dagger \hat P\hat S\right\rangle dW dW\nonumber\\
&&+\langle\hat S^\dagger \hat P+\hat P\hat S\rangle dW + O(dW dt). \label{eq-A}
\end{eqnarray}
The equation of motion for the ensemble-averaged expectation value
is obtained immediately from (\ref{eq-A}),
\begin{eqnarray}
\partial_t\overline{\langle\hat P\rangle}&=
&-i \overline{\langle [\hat P,\hat H] \rangle}\nonumber\\
&&-\frac12 \left( \overline{\langle\hat S^\dagger \hat S \hat
P\rangle+ \langle\hat P \hat S^ \dagger \hat S\rangle-2 \langle\hat
S^\dagger \hat P\hat S\rangle}\right)
\label{ope-dynamics}\\
&=&-i\left\langle\left[\overline{\hat P},\hat H\right]\right\rangle
\nonumber\\
&&-\frac12 \left( \langle\hat S^\dagger \hat S \overline{\hat
P}\rangle+ \langle\overline{\hat P} \hat S^\dagger \hat S\rangle-2
\langle\hat S^\dagger \overline{\hat P}\hat S\rangle\right).
\label{ope-dynamics-simple}
\end{eqnarray}
One could arrive at the same equation by using the master equation for the density matrix. It
is immediately seen that the coupling to an external environment, modifies the dynamics of the
observable if $\hat P$ and $\hat S$ do not commute, i.e., $[\hat P,\hat S]\not= 0$. The second
term in the right hand side of (\ref{ope-dynamics-simple}) describes the dynamics of the
expectation value in the presence of the relaxation induced by the environment. A steady state
for any observable will be reach when a competition between the two terms is established so
that to make the right hand side of (\ref{ope-dynamics-simple}) vanishing.

Of importance for the investigation of the dynamics of open quantum systems, is the time evolution of the single-particle density, i.e., the expectation value of the single-particle density operator. In standard quantum mechanics, i.e., for closed systems, the equation of motion for the single-particle density is the continuity equation which essentially embodies the important principle of particle conservation. Namely, if $n(r,t)$ is the single-particle density and $\vec j(r,t)$ the single-particle current density, we have
\begin{equation}
	\partial_t n(r,t)=-\nabla\cdot \vec j(r,t).
	\label{continuity-equation}
\end{equation}
When the system is open, the equation of motion for the single-particle density appears more complex. We have from (\ref{ope-dynamics-simple}) that
\begin{eqnarray} 
\partial_t \overline{n(r,t)}&= -\nabla\cdot \overline{\vec j(r,t)}-\frac12 \left( \langle\hat S^\dagger \hat S \overline{\hat n}\rangle+
\langle\overline{\hat n} \hat S^\dagger \hat S\rangle-2 \langle\hat S^\dagger
\overline{\hat n}\hat S\rangle\right). 
\label{continuity-equation-stochastic}
\end{eqnarray}
It may appear that this equation breaks the important physical principle of
particle conservation, since the second term in
(\ref{continuity-equation-stochastic}) can hardly be recognized as a
divergence of any current. Improbable as it might seem, Gebauer and
Car \cite{Gebauer2004a} were able to show that, at least in the Markov
approximation, the right hand side of
(\ref{continuity-equation-stochastic}) can be written as the divergence of
a new current density, $\vec{j}+\vec {j}_C$, where $\vec j_C$ describes the
transfer of momentum between the system and the environment.
                                            
It will be important when we will discuss in the following section the
development of a density functional approach to open quantum systems, so let
us discuss something that might appear as a trivial point. From a mathematical
point of view, the continuity equation, given the initial condition and the
current density $\vec{j}(r,t)$, uniquely determines the particle density everywhere and at
any time. This is because a simple integration of
(\ref{continuity-equation}) gives
\begin{equation}
	n(r,t)=n(r,t=0)-\int_0^t dt' \nabla\cdot\vec j(r,t').
\end{equation}

The opposite is not true. Given $n(r,t)$ and the initial conditions, the current density $\vec
j(r,t)$ is not uniquely determined. Indeed, given $\vec j(r,t)$ and $\vec j(r,t)+\nabla\times
\vec v(r,t)$, where $\vec v(r,t)$ is an arbitrary vector in space and time, these two currents
solve the continuity equation with the same density $n(r,t)$. For this reason, one usually
says that the particle density only determines the longitudinal current-density, while the
transversal part is left unknown. Indeed, only the longitudinal current is responsible for the
transfer of particle across any given surface.

The situation appears more complex if one starts with the continuity equation for open
systems. Even if one would consider the total current $\vec j+\vec j_C$, the density is not
uniquely determined. This is because the current $\vec j_C$ is determined by the density and
the current density themselves, therefore making (\ref{continuity-equation-stochastic}) a
non-linear equation whose solution appears non-trivial. The situation has not been clarified
and in the following we will assume that, similarly to what happens with the continuity
equation, (\ref{continuity-equation-stochastic}) uniquely determines the single-particle
density.
                                                                   
With logical steps similar to those who led us at the equation of motion for the single-particle density \cite{Giulianivignale}, we can derive the equation of motion for the ensemble-averaged current density for a system of interacting identical particles of mass $m$, in the presence of an external vector potential $\vec A_{ext}$,
\begin{eqnarray}
\partial_t \overline{\vec j(
r,t)}&=&\frac{\overline {n(r,t)}}{m} \partial_t {\vec A_{ext}}({
r},t) -\frac{\overline {{\vec j}({r},t)} }{m}\times\left[\nabla \times
{\vec A_{ext} }
({r},t)\right]\nonumber
\\&&+\frac{\langle\overline{\hat {\mathcal F}({
r},t)}\rangle}{m} +\langle\overline{\hat {\mathcal G}({r},t)}\rangle
\label{currenteq}
\end{eqnarray}
where we have defined
\begin{eqnarray}
\label{forces}
\hat {\mathcal G}({ r},t)&=\hat S^\dagger \hat j({ r},t) \hat
S  -\frac12 \hat j({ r},t) \hat S^\dagger \hat S
-\frac12 \hat S^\dagger \hat S \hat j({ r},t),\\
\hat {\mathcal F}({ r},t)&=-\sum_{i\not = j}\delta({ r}-\hat r_i)
\nabla_j U_{int}\left(\hat r_i-\hat r_j\right)+m \nabla
\cdot \hat \sigma({ r},t)
\end{eqnarray}
with the stress tensor $\hat \sigma({ r},t)$ given by
\begin{equation}
\hat \sigma({ r},t)=-\frac{1}{4}\sum_{i,j,k} \{\hat v_i,\{\hat
v_j,\delta({ r}-\hat r_k)\}\}.
\label{stresstensor}
\end{equation}
In these equations $\nabla_j$
contains the derivatives with respect to the coordinates of the
$j$-th particle, i.e., in 3D
$\nabla_j=(\partial_{x_j},\partial_{y_j},\partial_{z_j})$, and the current operator is defined as
\begin{equation}
\hat j(r,t)=\frac{1}{2}\sum_i\left\{\delta(r-\hat r_i), \hat
v_i\right\}
\end{equation}
with
\begin{equation}
\hat v_i=\frac{\hat p_i + e A_{ext}(\hat r_i,t)}m, \label{velocity}
\end{equation}
the velocity operator of particle $i$, and the symbol $\{\hat A,
\hat B\}\equiv (\hat A \hat B + \hat B \hat A)$ is the
anti-commutator of any two operators $\hat A$ and $\hat B$.             

The first two terms on the right hand side of (\ref{currenteq}) describe the
effect of the applied electromagnetic field on the dynamics of the
many-particle system; the third is due to particle-particle
interactions while the last one is the ``force'' density exerted by
the bath on the system. This last term is responsible for the
momentum transfer between the quantum-mechanical system and the
environment. It should be pointed out that the first three terms on the right hand side of (\ref{currenteq}) are present also in the standard equation of motion for the single-particle current density \cite{Giulianivignale}. Equation (\ref{currenteq}) can be seen as the equation for the ``forces'' acting on a single particle.

\subsection{Solving the Stochastic Schr\"odinger Equation}\label{solving_the_sse} 

For a single realization of the noise, the numerical solution of the SSE is similar to the
integration of the Schr\"odinger equation \cite{Kloeden1999,Higham2001}, and one can use some
of the standard techniques for the numerical integration of partial differential equations
\cite{Press1992}. However, the noise effectively reduces the stability and efficiency of the
numerical algorithms \cite{Kloeden1999}. For this reason, higher order techniques, like the
standard fourth order Runge-Kutta, do not offer the usual improvement as in the integration of
the standard dynamical equations. Indeed, the usual parameter used to identify this efficiency
is the scaling of the numerical error, $\epsilon$, in terms of the time step of the numerical
integration, $\Delta t$. Typically, we have a power law scaling, i.e.,
\begin{equation} 
	\epsilon\propto \Delta t^\nu.   
\end{equation}
For example, for ordinary differential equations, the second order Runge-Kutta has $\nu=2$,
while the fourth order Runge-Kutta has $\nu=4$. For Markovian systems, it is found quite
surprisingly that for any standard technique, e.g. the second and fourth order Runge-Kutta
methods, we have the same $\nu$ \cite{Rumelin1982}. Indeed, it has been shown that, in
general, any method which involves only Wiener processes, will have the same scaling
\cite{Rumelin1982} as the simpler Euler or Heun scheme (second order Runge-Kutta). High order
schemes which improve the scaling with the integration time step can be found, and have been
proposed --see for example \cite{Kloeden1999,Burrage1996} and references therein-- which
involve the evaluation of a class of stochastic processes, beyond the simple Wiener processes,
in building the numerical approximation to the exact solution of the SSE. These terms appear
in the Taylor-like expansion of the solution to high orders \cite{Kloeden1999,Burrage1996}. On
the other hand, to the best of our knowledge, for non-Markovian SSE there are not clear
results and a complete analysis of the error scaling is missing. The reader should also be
warned of some attempts at building high-order schemes to the solution of Markovian SSE, like
in~\cite{Wilkie2005} starting from the standard Runge-Kutta schemes. These are based on the
assumption $\Delta W\Delta W=\Delta t$, which is however valid only when the average over many
realizations of the noise is performed. Although it seems these schemes improves the
convergence, their validity is doubtful \cite{Burrage2006} and most likely restricted to a
small class of equations.

In this review, we will not discuss in details any algorithm of numerical solution of the SSE.
A nice step-by-step tutorial on the solution of stochastic differential equations can be found
in \cite{Higham2001}, while more advanced techniques can be found in
\cite{Kloeden1999,Burrage2006}. Here, we would like just to add that we can find a non-linear
SSE which preserves, at each time step, the norm of the wave-function and which reproduces the
same physical quantities as the linear SSE \cite{Ghirardi1990}. It is our experience that using this non-linear SSE usually improves the stability of the integration algorithm.

When dealing with the non-Markovian SSE, we face the problem of simulating the correlated
colored noise $\gamma(t)$ in (\ref{prop_noise}). Many solutions to this problem have been
proposed: they do differ on the amount of information we need at our disposal about the
function $C_{\alpha,\beta}(t)$. For example, for simple correlation functions the algorithm
proposed in \cite{Mannella1989} is probably the most efficient, but it does require the
knowledge of a first-order differential equation, whose $C_{\alpha,\beta}(t)$ is a solution.
This works well for $C_{\alpha,\beta}(t)\propto e^{-|t|}$ \cite{Mannella1989}. For more
complex cases, this approach can not be used, and we can revert to the solution proposed by
Rice \cite{Rice1944}, and recently revisited \cite{Billah1990}. This algorithm might not
be the most efficient \cite{Mannella1992, Billah1992}, but is based on the sole knowledge of
the Fourier Transform of $C_{\alpha,\beta}(t)$, the power spectrum, practically the minimal amount of information
we need. Moreover, this Fourier Transform can be known analytically for some models: this is
due to the fact that we might have access to, or we can approximate it to a certain degree,
the power spectrum of the bath. A variant of this algorithm has been recently proposed in
\cite{Barrat2011} for the case we know the function $C_{\alpha,\beta}(t)$ at each instance of
time, and we do not want to store the full Fourier Transform for computational purposes.
However, this algorithm does have a large numerical overhead that makes its application to the
SSE too expensive \cite{Biele2011}.

\subsection{Difference between the Stochastic Schr\"odinger equation and the 
master equation}\label{equivalence}                                                                                        
We want to discuss the equivalence between the master equation for the reduced density matrix
and the SSE. In particular, we want to show one case of general interest in which the two
formalisms are giving different results. We argue that this originates from the different ways
they deal with interaction.

Let us consider the Hamiltonian of a one dimensional boson system (in second quantization)
which, when the bath is not present, reads
\begin{eqnarray}
\hat H&=\hat H_0+\hat H^{int}=&\int
dx~\psi^\dagger(x)\left(-\frac{1}{2m}\frac{d^2}{dx^2}+V_{ext}(x)\right)
\psi(x)\nonumber\\
&&+\int dxdx'~ \psi^\dagger(x)\psi^\dagger(x')
U_{int}(x-x')\psi(x')\psi(x),
\end{eqnarray}
where $\psi(x)$ destroys a boson at position $x$, $V_{ext}(x)$ is a
confining potential, and $U_{int}(x-x')$ is the boson-boson
interaction potential. We consider here the case of a diluted gas of Alkali atoms, since this is a system of great interest \cite{Davis1995,Anderson1995,Dalfovo1999}. For these gases the interaction
potential can be substituted with the contact potential, i.e.,
\begin{equation}
U_{int}(x-x')=g_0(N-1)\delta(x-x')=\tilde{g}\delta(x-x')\label{Hint}
\end{equation}
where $g_0=4\pi\hbar^2 a/m$ is determined by the scattering length $a$ of the
boson-boson collision in the dilute gas and $N$ is the total number
of bosons in the trap, so that $||\psi||$=1 \cite{Dalfovo1999}.

In what corresponds to a Hartree approximation for the boson wave function, we
can go from the equation of motion for the annihilation operators to the
equation of motion for single-particle wave function $\Psi(x,t)$, when the external
bath is not coupled to the boson gas,
\begin{equation}
i\frac{d}{dt}
\Psi(x,t)=\left[-\frac{1}{2m}\frac{d^2}{dx^2}+V_{ext}(x)\right]\Psi(x,t)
+\tilde{g} n(x,t)\Psi(x,t)
\label{GP}
\end{equation}
where $n(x,t)=|\Psi(x,t)|^2$ is the single-particle density of the boson gas
\cite{Gross1963,Ginzburg1958}. Equation (\ref{GP}) (and its generalization to
two and three dimensions) correctly describes the physical properties
and especially the dynamics of a Bose-Einstein condensate
\cite{Davis1995,Anderson1995,Dalfovo1999}. That is, $\Psi(x,t)$ describes the dynamics of the ground state of the system, when the temperature of the gas is well below the critical temperature for the Bose-Einstein condensation \cite{Feynman,Huang2001,Dalfovo1999}.

A harmonic confinement is routinely generated in the magneto-optical traps used in the realization of the condensation in Alkali gases \cite{Davis1995,Anderson1995,Dalfovo1999}, so we choose
\begin{equation}
V_{ext}(x)=\frac12 m\omega_0^2 x^2.\label{harmonic}
\end{equation}

When the boson system is coupled to the external environment, we assume that
the Hamiltonian is not affected by this coupling, i.e., we neglect the Lamb shift, and the system evolves
according to the Markovian SSE
\begin{eqnarray}
d\Psi(x,t)&=&-i\left(-\frac{1}{2m}\frac{d^2}{dx^2}+\frac12 m\omega_0^2 x^2
+\tilde{g} n(x,t)\right)\Psi(x,t)dt\nonumber\\
&&-\frac12 \hat S^\dagger \hat S\Psi(x,t)dt
+\hat S\Psi(x,t)dW,
\label{sse-gp_notnorm}
\end{eqnarray}     
where we have assumed that the coupling with the environment is given by a single operator $\hat S$. Here, $dW$ is a standard Wiener process with properties (\ref{noise_ito}).

For convenience, we rescale this equation in terms of the physical quantities
$\omega_0$, $x_0=1/\sqrt{m\omega_0}$, and $g=\tilde{g}/x_0$ to arrive at
\begin{eqnarray}
d\Psi(x,t)&=&-i\omega_0\left(-\frac{x_0^2}{2}\frac{d^2}{dx^2}+\frac{x^2}{2 x_0^2}
+\frac{g}{\omega_0} n(x,t)x_0\right)\Psi(x,t)dt\nonumber\\
&&-\frac12 \hat S^\dagger \hat S\Psi(x,t)dt+\hat S\Psi(x,t)dW.
\label{sse-gp}
\end{eqnarray}

We begin with considering the case of non-interacting bosons, i.e., we
set $g = 0$. In this case, the Hamiltonian admits a natural complete
basis, the set of Hermite-Gauss wave-functions
\begin{equation}
\varphi_j (x)=\frac{1}{\sqrt{x_0 2^j j! \sqrt{\pi}}}\mathcal{H}_{j}(x/x_0)
e^{-x^2/2x_0^2},
\end{equation}
where $\mathcal{H}_j(x/x_0)$ are the standard Hermite polynomials \cite{Abramowitz1964}.
If we expand the wave-function
$\Psi(x,t)=\sum_{j}a_j(t)\varphi_j(x)$, and make use of the
orthonormality properties of the Hermite-Gauss wave functions, we
obtain the (stochastic) dynamical equation for the coefficients
$a_j$,
\begin{equation}
d a_i=\sum_j \left(H^0_{ij}a_j+\frac12 (\hat S^\dagger \hat
S)_{ij}a_j\right) dt+ dW \sum_j S_{ij}a_j
\label{freea}
\end{equation}
where $H^0_{ij}=(j+1/2)\omega_0 \delta_{ij}$.

Together with (\ref{freea}) we can study the dynamics of the
density matrix via the quantum master equation
(\ref{lindblad}), which in the same representation
as (\ref{freea}) reads
\begin{eqnarray}
\partial_t \rho_{ij}&=&-i\sum_k
\left(H^0_{ik}\rho_{kj}-\rho_{ik}H^0_{kj}\right)\nonumber\\
&&+\sum_{k,k'}\left(
S_{ik}\rho_{kk'} S^\dagger_{k'j}-\frac12 S^\dagger_{ik}
S_{kk'}\rho_{k'j}-\frac12 \rho_{ik} S^\dagger_{kk'}
S_{k'j}\right). \label{freerho}
\end{eqnarray}
The connection between (\ref{freea}) and (\ref{freerho}) is
established by the identity $\rho_{ij}=a^*_i a_j$ which is valid for any pair
of indexes $i$ and $j$. It was shown \cite{DAgosta2008a} that when the interaction is set to vanish, the two formalisms yield the same result in the limit of a large number of realizations.   

When one turns on the particle-particle interaction $U_{int}$, this
corresponds to adding to the free Hamiltonian an
interaction part, $\hat H^{int}$ that in the basis representation of the Gauss-Hermite
polynomials reads
\begin{equation}
H^{int}_{i,j}=\sum_{k,q}F_{i,j;k,q} a_k^* a_q
\end{equation}
where $F_{i,j;k,q}$ is the fourth-rank tensor defined as
\begin{eqnarray}
F_{i,j;k,q}&=&\frac{g}{\pi\omega_0}\frac{2^{-(i+j+k+q)/2}}{\sqrt{i!j!k!q!}}	\nonumber\\
	&&\times\int\limits_{-\infty}^\infty dx~ \mathcal{H}_i(x)\mathcal{H}_j(x)\mathcal{H}_k(x)\mathcal{H}_q(x)e^{-2x^2}.
\end{eqnarray}
A long but straightforward calculation, worked out in full detail in~\cite{Lord1949}, brings us to an expression for
$F_{i,j;k,q}$ in terms of Euler gamma functions and a hypergeometric
function~\cite{Lord1949,Abramowitz1964}. It can be shown that the hypergeometric
function reduces to the summation of a few -- at most $\min(i,j)$ --
terms \cite{DAgosta2008a}. In the case of the density matrix approach the interaction
Hamiltonian is immediately written as
\begin{equation}
H^{int}_{i,j}=\sum_{k,q}F_{i,j;k,q}\rho_{kq}.
\end{equation}

In solving the dynamics of the system described either by the
SSE~(\ref{sse-gp}) or the master equation~(\ref{freerho}) when the interaction is present, we have
assumed that at any instance of time the bath operator brings the
system towards the instantaneous ground state of the total {\it
interacting} Hamiltonian $H^0_{i,j}+H_{i,j}^{int}$. Formally, in the basis set that diagonalizes the total Hamiltonian, this corresponds to choosing
\begin{equation}
	S_{i,j}=\left(
	\begin{array}{ccccc}
	0 & 1 & 1 &\ldots &1\\
	0& 0& 0& \ldots& 0\\
	\vdots& \vdots& \vdots &\vdots &\vdots\\
	0& 0& 0& \ldots& 0
	\end{array}
	\right),
	\end{equation}
	for the bath operator. 
In addition, the
interaction potential (and hence the total Hamiltonian), being
defined in terms of the instantaneous density, is stochastic, namely
it is different for each element of the ensemble. While we
take this into account explicitly in the SSE~(\ref{sse-gp}), in the
master equation~(\ref{freerho}) we must consider the interaction
Hamiltonian averaged over all realizations. 

\begin{figure}[ht!]
\includegraphics[clip,width=8.5cm]{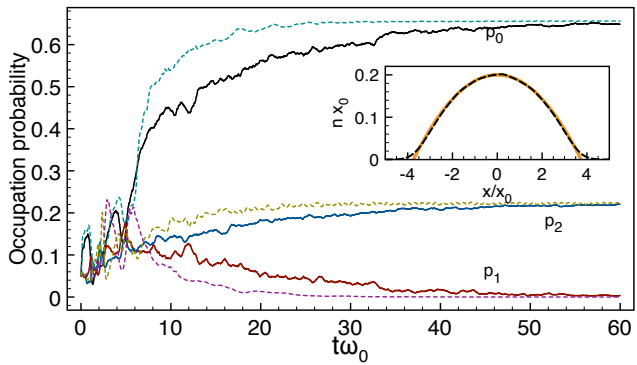}
\caption{
Probability of occupation of the first 3 energy levels of the interacting Hamiltonian versus
time calculated via the SSE (solid lines) averaged over 100 independent runs, and via the
equation of motion for the density matrix (dashed lines). In the inset we compare the
equilibrium density (black, dashed line) with the one obtained from the Thomas-Fermi
approximation to the ground state (orange, solid line). The initial condition is chosen by
putting the boson with equal probability in all the considered states.}
\label{projection}
\end{figure}

In figure~\ref{projection} we plot the occupation probability $p_j(t)$ of the state $j$ for
the first 3 energy levels of the free Hamiltonian ($p_j(t)=|a_j(t)|^2$ from the SSE or
$p_j(t)=\rho_{j,j}(t)$ from the density matrix). We have assumed an interaction strength
$g/\omega_0=5$, considered the first 20 energy levels, used a time step $\omega_0 \Delta
t=60/2^{17}$ and we have performed 100 independent runs of the SSE. The initial condition is
such that all the energy levels are occupied with the same probability. From the figure it is
evident that the system reaches, in the long time, the same steady state, but it is also clear
that the state calculated with the SSE relaxes slower to this steady state than the state
obtained from the density matrix equation. This is a spurious effect in the density matrix
dynamics where the average density defines the interaction potential without account for the
fluctuations of the state, and hence of the stochastic Hamiltonian.

We have also tested that the steady state reached during the dynamics is consistent with the
eigenstates of the Gross-Pitaevskii equation \cite{Dalfovo1999,DAgosta2002}. In particular,
the ground state of the interacting system, when the interaction is strong, can be obtained by
neglecting the kinetic contribution to the Hamiltonian. In this case, a good approximation to
the ground state density reads \cite{Dalfovo1999}
\begin{eqnarray}
|\Psi_0(x)|^2&=&\frac{\mu-1/2 m\omega_0^2 x^2}{gx_0}\theta \left(\mu-1/2 m\omega_0^2 x^2\right)\nonumber \\
&&+\mathrm{terms~proportional~to }~1/g^2,
\label{thomas-fermi}
\end{eqnarray}
where $\mu$ is the chemical potential, i.e., in this case the energy of the ground state, and
$\theta(x)=0$ if $x<0$ and $\theta(x)=1$ if $x\geq 0$.

In the inset of figure~\ref{projection} we plot the density obtained at $t\omega_0=60$ from
the SSE (black, dashed line) and the density obtained from the approximation
(\ref{thomas-fermi}) (orange, solid line). Notice that the value of the parameters $g$ and
$\mu$ have been obtained from the best fit with the numerics: indeed one can show that the
approximation (\ref{thomas-fermi}) is exact in the limit of very large interaction,
\cite{DAgosta2002,Dalfovo1999} which is not reached in our calculations.

We have therefore shown that the master equation and the SSE, for a system where the
Hamiltonian is stochastic, i.e., it does contain terms that depend on the wave function, could
lead to different results. This is expected since in the master equation we replace the
effective stochastic term with an ``average'' contribution, without having any control on it.
For example, in the case discussed the approximation corresponds to replacing terms like
$\overline{n(x,t)n(x',t)}$ with $\overline{n(x,t)}\times\overline{n(x',t)}$. In the SSE, this
approximation is not performed, since the average over the stochastic noise is done after the
full evolution is calculated.


\section{Stochastic Density Functional theory}\label{stochastic_dft} 

As we will see for example in section \ref{spin_thermal}, one severe limitation of the
applicability of the theory is the presence of interaction between the particles of the
system. This is nothing new since similar problems arise also in the case of closed systems
where the presence of interaction makes obtaining the exact solution of the dynamics of the
system an almost impractical task. However, in the theory of many-particle systems, whenever
we are only interested in certain quantities, it has been found that the complexity of the
problem can be greatly reduced, up to the level of making it treatable with numerical
techniques. These results go under the name of Density Functional Theory
(DFT)~\cite{Hohenberg1964,Kohn1965,Marques2006}. Nowadays, there are different flavors of what
one can call Density Functional Theories, and the interested reader can find better
introductions to that field for example in \cite{Marques2006}. For our purposes, it is
sufficient to say that, if we focus on the dynamics of one observable, let us say the single
particle density, then we can obtain this dynamics by investigating a system of
non-interacting particles, called Kohn-Sham (KS) system in the presence of a certain external
potential \cite{Runge1984,vanLeeuwen1999}. This potential is built to give \emph{exactly} the
dynamics of the single-particle density of the real system~\cite{Marques2006}. However, it is
important to stress that any other quantity calculated within the KS system and which cannot
be expressed in terms of the single-particle density alone, cannot, in general, be trusted.

For the sake of this review, of even more interest is the case in which the current density is
the quantity we would like to obtain. For this case, it has been shown that one can build a
system of non-interacting quasi particles that, in the presence of an appropriate vector
potential, is able to mimic the dynamics of the current density of the real system
\cite{Vignale2004,Vignale2008}. Again, one must add the usual caveat that in principle only
the current density and all the physical quantity that can extracted from it with simple
operations (like for example the single-particle density) have a physical meaning. All the
others bear little resemblance to the same quantity of the real system, although, in some
cases one can realize that some physical information can still be extracted from the dynamics
of the Kohn-Sham system \cite{DAgosta2005a}.

The advantage of this formulation is manyfold and so far reaching that W. Kohn was awarded the
Nobel Prize in Chemistry for the initial formulation of the theory which was dealing with the
properties of the ground state. Indeed, it essentially paves the way to the numerical
investigation of complex many-body systems. It would be of great interest if one were able to
bring together the formulation of the open quantum systems we have discussed so far and the KS
theory of many-body systems. This result has been achieved recently by various groups
\cite{Burke2005,DiVentra2007,DAgosta2008a,Yuen-Zhou2009a} with different formulations of the
theory. Also in this case, the parallel between the density matrix formalism and the SSE
appears evident and not surprisingly in certain cases one can reproduce the result of one
formalism into the other. However, as we will point out more clearly in the following, a KS
formulation in terms of the density matrix of open quantum systems appears flawed
\cite{DAgosta2008a} since at the very basic level the KS Hamiltonian does contain non-linear
terms of the state of the system thus making the derivation of a closed equation of motion for
the density matrix problematic as we have discussed in section \ref{open_quantum_systems}.
                  
\subsection{Stochastic Time-Dependent (Current-)Density Functional Theory}

Let us begin by revising, without proof, the theorems of time-dependent density functional
theory. We consider $N$ non-relativistic particles of charge $e$, mutually interacting via the
Coulomb interaction, in a time-dependent external potential, either a vector or scalar
potential. We consider the general Hamiltonian
\begin{equation}
\hat H=\sum_{i=1}^N \frac{\left[\hat p_i +e \vec A_{ext}(\hat
r_i,t)\right]^2}{2m}+V_{ext}(\hat r_i,t)+\sum_{i < j}^N U(\hat r_i-\hat r_j)
\label{h-manybody}
\end{equation}
where $\vec A_{ext}(r,t)$ is the external vector potential, $V_{ext}(r,t)$ the external scalar potential, and
$U(r)$ is the particle-particle interaction
potential. The many-body system evolves according to the time-dependent Schr\"odinger equation,
\begin{equation} 
	i\frac{d|\Psi(t)\rangle}{dt}=\hat H |\Psi(t)\rangle.
\end{equation} 

When there is only a scalar potential $V_{ext}$, we have the following theorems of Time-Dependent Density Functional Theory (TDDFT)\cite{Runge1984,vanLeeuwen1999,Vignale2004,Ruggenthaler2011,Marques2006}:
\begin{theo}[Runge-Gross \cite{Runge1984}]
    We consider $N$ non-relativistic electrons, mutually interacting via the
   Coulomb repulsion, in a time-dependent external potential. Two single-particle densities
   $n(r,t)$ and $n'(r,t)$ evolving from a common initial state $|\Psi(t =
   0)\rangle$ under the influence of two potentials $V_{ext}(r,t)$ and
   $V_{ext}'(r,t)$ (both Taylor expandable about the initial time 0)
   eventually differ if the potentials differ by more than a purely
   time-dependent (r-independent) function: $V_{ext}(r,t)-V'_{ext}(r,t) \not=
   c(t)$. Under these conditions, there is a one-to-one mapping between
   densities and potentials, which implies that the potential is a functional
   of the density.
\end{theo}         
The theorem is similar to the one that Hohenberg and Kohn \cite{Hohenberg1964}
proved in 1964 and that allowed, a year later, Kohn and Sham \cite{Kohn1965} to
formulate the DFT. The key point being that one can
obtain physical information about a many-body system by investigating the
behaviour of a simpler system where particle-particle interaction is turned
off.

The initial formulation of the Runge-Gross theorem was not satisfactory since
it was based on the existence of some action that in the dynamical case did
not respect causality \cite{vanLeeuwen1999}. Some years later, van Leeuwen was
able to extend the theorem and prove that there is no need to define an
action from where the equation of motion can be derived. Recently, Vignale has
shown that the initial formulation can be made consistent by considering carefully
the boundary and initial conditions for the dynamics
\cite{Vignale2008}.
\begin{theo}[van Leeuwen \cite{vanLeeuwen1999}]
Let $\hat H(t)$ and $\hat H'(t)$ be two Hamiltonians of the form (\ref{h-manybody})
containing not only two different time-dependent local potentials
$V_{ext}(r,t)$ and $V'_{ext}(r,t)$ but also two different particle-particle
interactions $U$ and $U'$. Let $n(r,t)$ be the density that evolves from the
initial state $|\Psi(t=0)\rangle$ under the Hamiltonian $\hat H(t)$, and let
$|\Psi'(t=0)\rangle$ be another initial state with the same density and the
same value of the density gradient. Then the time-dependent density $n(r,t)$
uniquely determines, up to a time-dependent constant, the potential
$V_{ext}'(r,t)$ that yields $n(r,t)$ starting from $|\Psi'(t=0)\rangle$ and
evolving under $\hat H'$.
\end{theo}                                    
It can be easily seen that the van Leeuwen's theorem include the Runge-Gross results as a simple corollary, by considering two systems evolving with the same interaction and with two external local potentials.  

The case for which a vector potential is present has been investigated first by Ghosh and Dhara \cite{Ghosh1988} and later Vignale proposed a theorem along the proof of van Leeuwen. Here for simplicity we report only the second theorem, again the result by Ghosh and Dhara follows as a simple corollary. Vignale's theorem for Time-Dependent Current-Density Functional Theory (TDCDFT) states:

\begin{theo}[Vignale \cite{Vignale2004}]
	Consider a many-particle system described by
	the time-dependent Hamiltonian 
	\begin{equation}
	\hat H(t) =\sum_i\left[\frac{1}{2m}\left( \hat p_i+\vec A_{ext}(\hat r_i,t)\right)^2+V_{ext}(\hat r_i,t)\right]+\sum_{i<j}U(\hat r_i-\hat r_j), 
\end{equation} 
where $V_{ext}(r,t)$ and $\vec A_{ext}(r,t)$ are given external scalar and vector potentials,
which are analytic functions of time in a neighborhood of $t=0$, and $U(r_i,r_j)$ is a
translationally invariant two-particle interaction. Let $n(r,t)$ and $\vec j(r,t)$ be the
particle density and current density that evolve under $\hat H$ from a given initial state
$|\Psi(0)\rangle$. Then, under reasonable assumptions, the same density and current density
can be obtained from another many-particle system, with Hamiltonian
\begin{equation}
\hat H'(t) =\sum_i\left[\frac{1}{2m}\left( \hat p_i+\vec A'(\hat r_i,t)\right)^2+V'(\hat r_i,t)\right]+\sum_{i<j}U'(\hat r_i-\hat r_j),
\end{equation}                                 
starting from an initial state $|\Psi'(0)\rangle$ which yields the same
density and current density as $|\Psi(0)\rangle$ at time $t = 0$. The potentials $V'(r,t)$ and $\vec A'(r,t)$ are uniquely determined by $V_{ext}(r,t)$ and $\vec A_{ext}(r,t)$, $|\Psi(0)\rangle$, and $|\Psi'(0)\rangle$ , up to gauge transformations of the form 
\begin{eqnarray}
	V'(r,t)\rightarrow V'(r,t)-\frac{\partial \Lambda(r,t)}{\partial t},\nonumber\\
	\vec A'(r,t)\rightarrow \vec A'(r,t)+\nabla \Lambda(r,t),
\end{eqnarray} 
where $\Lambda$ is an arbitrary regular function of $r$ and $t$, which satisfies the initial condition $\Lambda(t=0)=0$.
\end{theo}
                
TDDFT and TDCDFT have been quite successful in describing the dynamics of closed quantum
systems, especially whenever one needed to calculate the response of the system to an external
perturbation \cite{Marques2006}. These theories allow to go beyond the actual state-of-the-art
in many fields, e.g., in electron transport in nanoscale devices where it does predict novel
and interesting results \cite{Sai2005}. The secret of this success is in the fact that the
theory allows for treating complex system with relative ease, since only the dynamics of a
closed non-interacting many-body system is computed.

It appears therefore natural to try to generalize the theorem of TDDFT and TDCDFT to the case
of open quantum systems. This generalization has been achieved by using both the reduced
density matrix and the SSE formalisms \cite{Burke2005, DiVentra2007, DAgosta2008a}. Let us
assume that the quantum-mechanical system described by the Hamiltonian (\ref{h-manybody}) is
coupled, via given many-body operators, to one or many external baths that can exchange
energy and momentum with the system. If we assume that the dynamics of each bath is
described by a series of independent memory-less processes, the dynamics of the system is
governed by the stochastic Schr\"odinger equation in the Markov approximation,
\begin{eqnarray}
d |\Psi(t)\rangle&=&-i \hat H |\Psi(t)\rangle dt-
\frac12\sum_{\alpha}
\hat S^\dagger_\alpha \hat S_\alpha |\Psi(t)
\rangle dt\nonumber\\
&&+\sum_\alpha dW_\alpha(t) \hat S_\alpha |\Psi(t)\rangle
\label{stochasticse}
\end{eqnarray}
where $\{\hat S_\alpha\}$ describe the coupling of the system with the environment. In
\ref{stochasticse}, we could have time-dependent Hamiltonian and bath operators. This would
not change the following discussion.

As discussed in section \ref{observable-equation}, we here assume that the knowledge of the
current density is sufficient to obtain the single-particle density, and moreover this
solution is unique, i.e., it depends solely on the single-particle current density and the
initial conditions. If this is not the case, the proof of the following theorem, as we report
it, does not hold and we have to revert to a more cumbersome yet similar proof where we need
to assume that the ensemble-averaged single-particle density is differentiable at all order in
time.

We have the following result \cite{DiVentra2007}:
\begin{theo} 
    Consider a many-particle system described by the dynamics in (\ref{stochasticse}) with the
many-body Hamiltonian given by (\ref{h-manybody}). Let $\overline {n({ r},t)}$ and $\overline
{{\vec j}({ r},t)}$ be the ensemble-averaged single-particle density and current density,
respectively, with dynamics determined by the external vector potential ${\vec A_{ext}}({
r},t)$ and bath operators $\{\hat S_{\alpha}\}$. Under reasonable physical assumptions, given
an initial condition $|\Psi(0)\rangle$ and the bath operators $\{\hat S_{\alpha}\}$, another
potential $\vec A'({ r},t)$ which gives the same ensemble-averaged current density must
necessarily coincide, up to a gauge transformation, with $\vec A_{ext}({r},t)$.
\end{theo}   

A theorem of similar content for the one-to-one correspondence between the single-particle
density and the scalar potential has been discussed in ~\cite{Burke2005}. The starting point
of those Authors has been the equation of motion for the density matrix of the system,
described in the Markov approximation. In \cite{Yuen-Zhou2009a, Yuen-Zhou2010} the theorem by
Burke et al. \cite{Burke2005} and the previous one have been extended to the non-Markovian
dynamics. The proofs of these theorems follow essentially the same logical steps as the ones
we will present in the following.

{\it Proof:} We follow a line of reasoning common to the proofs of the van
Leeuwen and Vignale theorems \cite{vanLeeuwen1999,Vignale2004}. Recently, a more elegant proof of Runge-Gross and van Leeuwen's theorems has been proposed \cite{Ruggenthaler2011}, but the application of this method to the equation of motion for the current-density appears not clear. Let us continue by assuming
that the current density $\overline {{\vec j}({r},t)}$ is obtained from
another many-particle system with Hamiltonian
\begin{equation}
\hat H'(t)=\sum_{i} \frac{\left[\hat p_i+e
    {\vec A}'(\hat r_i,t)\right]^2}{2m}+\frac12 \sum_{i\not = j}U'\left(\hat r_i-\hat
r_j\right), \label{h1}
\end{equation}
evolving from an initial state $|\Psi'(0)\rangle$ and following the stochastic Schr\"{o}dinger
equation~(\ref{stochasticse}) with the {\em same} bath operators $\hat S_\alpha$. By
assumption, $|\Psi'(0)\rangle$ gives --in the primed system-- the same initial current and
particle densities as in the unprimed system. The equation of motion for the ensemble-averaged
current density is (\ref{currenteq}).
                   
Equations similar to (\ref{stresstensor})--(\ref{velocity}) are written for the system with
the vector potential ${\vec A}'({r},t)$. Similar force terms $\mathcal{F}'$ and $\mathcal{G}'$
appear in these equations. $\mathcal{F}'$ and $\mathcal{G}'$ differ from the forces in the
unprimed system, since the initial state, the external vector potentials, and the velocity
$\hat v$ are different. By assumption, the ensemble-averaged current and particle densities
are the same, therefore
\begin{eqnarray}
\partial_t \overline {{\vec j}({ r},t)}&=&\frac{\overline {n({r},t)}}{m}
\partial_t {\vec A}'({ r},t) -\frac{\overline {{\vec j}({ r},t)}
}{m}\times\left[\nabla \times {\vec A}'({r},t)\right]\nonumber\\
&&\label{currentq1}+\frac{\langle\overline{\hat {\mathcal F}'({
r},t)}\rangle}{m} +\langle\overline{\hat {\mathcal G}'({
r},t)}\rangle\label{currentqp}.
\end{eqnarray}
Taking the difference of (\ref{currenteq}) and
(\ref{currentqp}) we arrive at
\begin{equation}
\overline {n({r},t)} \partial_t \Delta \vec A(r,t) =\overline {{\vec
j}({r},t)} \times\left[\nabla \times \Delta {\vec A}({
r},t)\right]+\Delta Q({r},t) \label{difference}
\end{equation}
where $\Delta {\vec A}({r},t)\equiv {\vec A}'({r},t)-{\vec
A_{ext}}({r},t)$ and $\Delta Q({r},t)\equiv Q'({r},t)-Q({
r},t)$ with
$Q({r},t)=\langle\overline{\hat {\mathcal F}({
r},t)}\rangle+m\langle\overline{\hat {\mathcal G}({r},t)}\rangle,$
and $Q'({r},t)$ the same quantity but in the primed system.

We next prove that (\ref{difference}) admits only one
solution, i.e., $\Delta {\vec A}({r},t)$ is completely determined
by the averaged dynamics of the current and particle densities, once
the coupling with the environment via $\hat S_\alpha$, is assigned.
To this end we expand (\ref{difference})
 in series about $t=0$ and obtain an equation for the
$l$-th derivative of the vector potential $\Delta {\vec A}({
r},t)$. We thus arrive at the equation
\begin{eqnarray}
\overline {n_0({r})}(l+1)\Delta \vec A_{l+1}({r})=&-\sum_{k=0}^{l-1}(k+1)\overline {n_{l-k}({r})}\Delta \vec A_{k+1}({r})\nonumber\\
&+\Delta Q_l({r})\nonumber\\
&+\sum_{k=0}^l \overline {\vec j_{l-k}({r})}\times \left[\nabla\times
\Delta \vec A_{k}({r})\right] \label{expansion}
\end{eqnarray}
where, given an arbitrary function of time $f({r},t)$, we have
defined the series expansion
\begin{equation}
f_l({r})\equiv \frac{1}{l!}\left.\frac{\partial^l f({r},t)}{\partial t^l}\right|_{t=0}.
\end{equation}
To prove that the right hand side of (\ref{expansion}) does not contain any term
$\Delta \vec A_{l+1}({r})$ we use the fact that the dynamics of any
ensemble-averaged operator is given by (\ref{ope-dynamics-simple}).
Indeed, this implies that the $l$-th time derivative of any operator can be
expressed in terms of its derivatives of order $k<l$, time derivatives of the
Hamiltonian of order $k<l$, and powers of the operators $\hat S_\alpha$ and $\hat
S_\alpha^\dagger$. The time derivatives of the Hamiltonian do contain time
derivatives of the vector potential ${\vec A}({r},t)$, but always of order
$k<l$. Then on the right-hand side of (\ref{expansion}) no time derivative
of order $l+1$ appears. Equation~(\ref{expansion}) can be thus viewed as a
recursive relation for the time derivatives of the vector potential $\Delta
{\vec A}({r},t)$. To complete the recursion procedure we only need to
assign the initial value $\Delta {\vec A}_0({r})={\vec A'}({r},t=0) -
{\vec A}_{ext}({r},t=0)$. Since in the unprimed and primed systems the densities
and current densities are, by hypothesis, equal, the initial condition is
simply given by $\overline {n({r},t=0)}\Delta A_0({r})=\langle
\Psi(0)|\overline{\hat j_p (r,t=0)}|\Psi(0)\rangle-\langle
\Psi'(0)|\overline{\hat j_p (r,t=0)}|\Psi'(0) \rangle$, where $\hat j_p({
r})=(1/2m)\sum_i \{\hat p_i,\delta({r}-\hat r_i)\}$ is the paramagnetic
current density operator.

The same considerations as in~\cite{Vignale2004} about the finiteness of the convergence
radius of the time series~(\ref{expansion}) apply to our case as well. We rule out the case of
a vanishing convergence radius by observing that it seems implausible that the smooth (in the
ensemble-averaged sense) dynamics induced by (\ref{stochasticse}) can introduce a dramatic
explosion of the initial derivatives of $\Delta {\vec A}({r,t})$. If this holds, the expansion
procedure~(\ref{expansion}) can be iterated from the convergence radius time onward. We have
then proved that (\ref{expansion}) completely determines the vector potential $\Delta {\vec
A}({r},t)$ and thus, since ${\vec A_{ext}}({r},t)$ is assumed known, it determines ${\vec
A}'({r},t)$ uniquely, up to a gauge transformation.

To finalize our proof, we consider the case in which $U=U'$ and
$|\Psi(0)\rangle=|\Psi'(0)\rangle$. If this holds, $\Delta {\vec A}_0({r})\equiv 0$. Then the
recursion relation admits the unique solution $\Delta {\vec A}_l({r})\equiv 0$ for any $l$,
and at any instant of time $t$ we have ${\vec A({r},t)}={\vec A'({r},t)}$ (still up to a gauge
transformation).\qed
                          
The theorem states that the one-to-one correspondence is between the averaged current density and the vector potential applied to the system. In view of this, the KS vector potential is therefore a functional of the averaged current density, i.e.,
$\vec A_{KS}(r,t)=\vec A'[\overline{\vec j(r,t)},|\Psi(0)\rangle](r,t)$. Obviously, as in the case of any DFT theory, nobody knows the form of this functional, therefore the quality of the results we will obtain from our stochastic approach depends on the quality of the approximations we are able to make for this functionals. At the moment, this is an open area of research. The hope is obviously that standard approximations that have proven very useful in the past, like for example the adiabatic-local-density approximation \cite{Marques2006,Giulianivignale} and its refinements, provides a solid foundation to explore the reach of this approach.   

By looking at the statements of the theorems, we notice that they all establish a one-to-one correspondence between two quantities that have the same ``spatial dimension''. With that we mean that the correspondence is either between a scalar (the single-particle density) and another scalar (the potential) or between a vector (the single-particle current density) and another vector (the vector potential). It is then easy to prove that if one tries to establish a correspondence between the current density and the scalar potential, is going to fail \cite{DAgosta2005a}. From a mathematical point of view this appears quite easy to understand: in creating a connection between two different spaces (the one of the densities and the one of the potentials, for examples) one has to be sure that the dimensions of these spaces are the same. For this reason, one should be able to accept that, if the continuity equation is not valid (see discussion in section \ref{observable-equation}), then the only physical information we can extract by using the TDCDFT theorem is the current density. For this reason, for example, the results of \cite{Yuen-Zhou2009a} appear lacking a solid foundation.                                 
                    
\subsection{Closing the system}
The proof we have presented for the theorem 4 leaves open a very fascinating possibility. What
happens if we do assume that the bath operators $\hat S_\alpha$ are not the same in the two
systems? Interestingly, it can be seen that the proof remains the same, i.e., one arrives at
the conclusion that whenever the interaction potentials, initial conditions, and now the bath
operators are the same then the vector potential is uniquely identified
\cite{Yuen-Zhou2009a}. A similar result holds for the scalar potential \cite{Yuen-Zhou2010}.
One therefore arrives at the amazing conclusion that, if one is interested in the dynamics of
selected observables then we can close the system, i.e., we set $\hat S_\alpha\equiv 0$ in the
KS system, and still get the exact result. The KS system becomes then a closed and isolated
system, exactly identical to the ones we already know in standard quantum mechanics. We
therefore do not need to study the evolution of the system under a random influence of the
environment. Although this possibility
opens up a new way to look at the dynamics of open quantum systems, care needs to be applied
in understanding its range of applicability. Indeed, one can easily see that the KS scalar or
vector potentials needed to reproduce the exact dynamics are functional of the coupling with
the external environment. A priori, very little is known on the way this coupling enters in
the KS potentials, and it is very likely that any local approximation, so successful in DFT,
will fail miserably. Moreover, the KS potential in this way loses completely its generality,
since it will be strongly affected by the system under investigation. It is clear, however,
that further investigation along these very interesting and potentially promising lines is
needed.

\section{Applications of the Stochastic Schr\"odinger
Equation}\label{applications}

In this section, we will discuss some applications of the stochastic Sch\"odinger equation to
investigate the dynamics of interesting systems. Our excursion must be limited, due to the
variety of approaches that have characterized this field. Our point of view will be starting
from systems that evolve according to a SSE as we have discussed in section
\ref{sse}. This cuts out a certain number of interesting, phenomenological, results that do
not fit easily in this approach. In selecting the material for this review, we have indeed
preferred to maintain a certain degree of consistency, rather than present a scattered amount
of results. The interested reader can look into a series of beautifully written books where
these topics are covered in more details
\cite{Gardiner2000,Gardiner2004,Razavy2006,Weiss2007}.

In selecting the topics, we have also maintained a focus on solid state physics to match the
audience of this journal: we will discuss spin thermal transport, the onset of Bose-Einstein
condensation in Alkali gases, the general problem of relaxation in the presence of a thermal
bath at given temperature. This list cannot be exhaustive of the field but, we believe, it
gives a flavour of the directions of research one could tackle within this working framework.


\subsection{Spin Thermal Transport}\label{spin_thermal} 
\begin{figure}[ht!]
	\includegraphics[width=10cm]{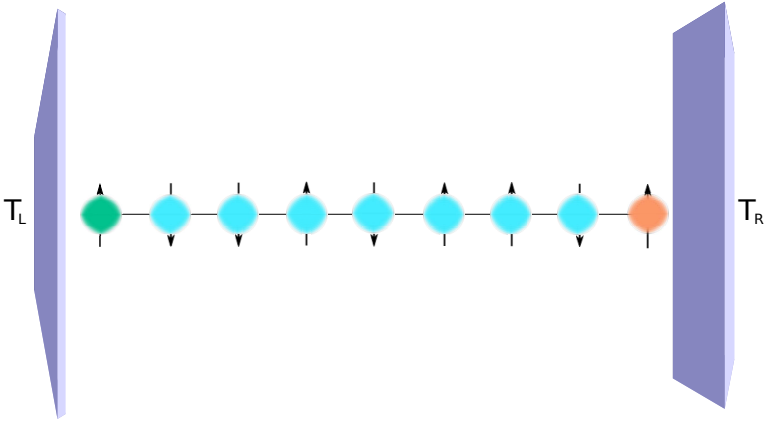}
	\caption{A chain of spins is connected to two energy reservoirs having different temperatures via the spins at the beginning and end of the chain. The reservoirs transfer or absorb energy from the chain. The spins are interacting with a nearest neighborhood interaction of parameter $Q$. A position-independent magnetic field $\vec h=\omega/2 \hat z$ is present. }
	\label{spin_chain}
\end{figure}

A neat application of a SSE is the investigation of the thermal transport in one dimensional
spin chains. The simplest system we can think of is a chain of $N$ spin $1/2$ atoms kept at
fixed positions. A nearest neighborhood interaction is present, and we can consider also the
presence of a magnetic field. The two ends of the chain are connected to two thermal
reservoirs at different temperatures: those two spins fluctuate due to the stochastic
interaction with the reservoir, energy is transferred or absorbed from the neighboring sites,
and the whole chain reaches a steady state in the long time regime. A simple schematic of the
system can be seen in figure~\ref{spin_chain}.
The Hamiltonian for this system, when the coupling is not present, is written as 
\begin{equation}
	\hat H=-Q\sum_{n=0}^{N-2}\vec \sigma_n \cdot \vec\sigma_{n+1}+\sum_{n=0}^{N-1}\frac{\omega}2 \sigma^z_n
\end{equation}
where, $\vec \sigma_n=\left(\sigma_n^x,\sigma_n^y,\sigma_n^z\right)$, $\sigma^i_n$ is the spin operator of the site $n$ along the direction $i=\{x,y,z\}$, and $Q$ is the interaction constant we have assumed constant through all the sites. We have embedded the system into a uniform magnetic field which we assume oriented along the $\hat z$ axis and of strength $\omega/2$. For a $N$ site chain we have that 
\begin{equation}
	\sigma_n^i=\mathop{\underbrace{\eins_{}\otimes \eins_{}}}_{n} \otimes \hat s^i \otimes \mathop{\underbrace{ \eins_{}\otimes \eins_{}\otimes\eins_{}}}_{N-n-1},\label{spin}
\end{equation}                                                                                            
where $\eins_{}$ is the identity in the $2\times2$ space and $\hat s^i$ is the $i$-th $2\times2$ Pauli matrix. It is then clear that each operator $\sigma_n^i$ is a sparse matrix of dimension $2^N\times 2^N$. 

An effective way to define the effect of the thermal reservoirs is the following. At each instance of time, the direction
of the first and last spin are randomized: the new spin directions are chosen according to a Boltzmann statistical weight
which depends on the ``local temperature'' of the last and first spins, i.e., to the temperature of the reservoir they
are attached to \cite{Mejia-Monasterio2007}. An equivalent way is to solve the dynamics via either a Lindblad equation
or, more conveniently, via an appropriate SSE \cite{Mejia-Monasterio2007,Wichterich2007}. The Lindblad equation for the system is written as
\begin{equation}
	\frac{d} {dt}\hat \rho(t)=-i\left[\hat H,\hat \rho(t)\right]+\mathcal{D}_L\hat \rho(t)+\mathcal{D}_R\hat \rho(t)
\end{equation}
where
\begin{eqnarray}
	\mathcal{D}_L\hat \rho(t)=\sum_{k,l=1}^2\gamma_{k,l}(T_L)\left(F_k \hat \rho(t) F_l^{\dagger}-\frac12 F_k^{\dagger} F_l\hat \rho(t)-\frac12\hat \rho(t)F_k^{\dagger} F_l\right),\\
	\mathcal{D}_R\hat\rho(t)=\sum_{k,l=1}^2\gamma_{k,l}(T_R)\left(G_k \hat\rho(t) G_l^{\dagger}-\frac12 G_k^{\dagger} G_l\hat \rho(t)-\frac12 \hat\rho(t)G_k^{\dagger} G_l\right).
	\label{bath_coupling}
\end{eqnarray}          
In (\ref{bath_coupling}), the operators $F$ and $G$ are the raising and lowering operators for the first and last spin, respectively, i.e.,
\begin{equation}
	F_1=\sigma_0^+,~F_2=\sigma_0^-,~G_1=\sigma_{N-1}^+,~G_2=\sigma_{N-1}^-,
\end{equation}                                                      
which are build starting from the raising and lowering spin $1/2$ $2\times 2$ operators in a similar fashion as in (\ref{spin})
\begin{equation}
	\hat s^+=\hat s^x+i\hat s^y,~~\hat s^-=\hat s^x-i\hat s^y.
\end{equation}
The coupling matrix ${\bf \gamma}$ is given by \cite{Wichterich2007}
\begin{equation}
	\gamma(\omega)=\lambda I(\omega)\left(
	 \begin{array}{cc}
		f(\beta,\omega) & \sqrt{f(\beta,\omega)^2+f(\beta,\omega)}\\ 
		\sqrt{f(\beta,\omega)^2+f(\beta,\omega)} & f(\beta,\omega)+1
	 \end{array}
	\right)
\end{equation}
where $f(\beta,\omega)=\left[\exp{(\beta \omega})-1\right]^{-1}$ is the Bose-Einstein distribution function at temperature $\beta=1/k_B T$, $I(\omega)$ is the bath spectral function, and $\lambda$ is the coupling strength, all evaluated at the energy of the magnetic field, $\omega$. Once the matrix $\gamma$ is diagonalized with $\alpha_1=0$ and $\alpha_2$ as the eigenvalues, one can rewrite the Lindblad equation in the equivalent form
\begin{eqnarray}
	   \frac{d} {dt}\hat \rho(t)&=&-i\left[\hat H,\hat \rho(t)\right]\nonumber \\
	&&+\alpha_{2}(T_L)\left(L^L \hat \rho(t) {L^L}^{\dagger}-\frac12 {L^L}^{\dagger} L^L\hat \rho(t)-\frac12 \hat \rho(t){L^L}^{\dagger} L^L\right)\nonumber\\
	&&+\alpha_{2}(T_R)\left(L^R \hat \rho(t) {L^R}^{\dagger}-\frac12 {L^R}^{\dagger} L^R\hat \rho(t)-\frac12 \hat \rho(t){L^R}^{\dagger} L^R\right),
\end{eqnarray}
where the operators $L^L$ and $L^R$ are linear combination of the operators $F$ and $G$.
Being in the diagonal Lindblad form, we know that we can easily rewrite the dynamics induced by this master equation into an equation for a stochastic wavefunction $\Psi(t)$ \cite{Mejia-Monasterio2007}. For this problem, going from the master equation to the SSE offers a great numerical advantage. The dimension of each of the matrices entering the Lindblad equation is indeed $2^N\times 2^N$, which even for a short chain of $10$ spins, means more than $10^6$ elements. Although many of those matrices will be sparse, the same is not true for the density matrix itself, therefore one cannot rely on using any efficient numerical tool for sparse matrix. In contrast, for the SSE the wavefunction is a vector of $2^N$ elements and since all the spin operators are sparse matrices, one can reduce the amount of memory and operations by order of magnitudes, thus making a larger number of spins in the chain affordable. Even in this case, the number of elements in the chain has to be restricted to about 20. Two ways out are possible. On the one hand, a reformulation of the problem in terms of Majorana fermions \cite{Prosen2008} seems to allow for longer chains, up to 100 elements, to be tackled at least to investigate the steady-state regime. On the other hand, one could reformulate the problem in terms of spin-density waves \cite{Giulianivignale}: by reducing the Hilbert space to the low energy waves, one could effectively reduce the dimensionality of the problem.  
      
\begin{figure} 
	\includegraphics[width=10cm]{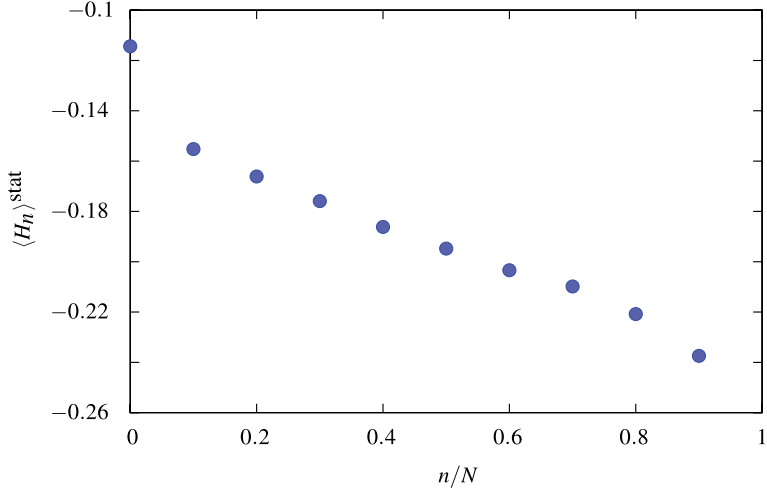}
	\caption{Ergodic average of the magnetic energy of each site, $H_n$, over a long-time single realization of the stochastic noise. The parameters $\beta_L= 0.41$, $\beta_R = 1.39$, $Q=\lambda=0.01$, and $\omega = 1$ have been chosen. For the realization a final time $t_{fin}=10^5$ in units of $\omega^{-1}$ has been used. From \cite{Mejia-Monasterio2007}, reprinted with permission.}
	\label{mj4}
\end{figure}

In figure \ref{mj4}, the magnetic energy of a given site is plotted, 
\begin{equation}
	\langle H_n\rangle^{stat}=\frac{1}{T+1}\sum_{k=0}^T\langle \Psi(t_k) |H_n |\Psi(t_k) \rangle,
	\label{stat_aver}
\end{equation}    
where $H_n=\omega\sigma^z_n/2$ is the magnetic energy of site $n$, $t_k=k\Delta t$ is the discrete-time grid of the simulation, and the average in (\ref{stat_aver}) is obtained from a single realization of the random noise using the ergodic theorem where we replace the average over many realizations of the noise with the time average of a long time realization. Here, $T$ is the number of time steps used for the ergodic average after the steady state has been reached. In this simulation, $T=10^5$.  
With a thermal gradient, there is an energy imbalance between the left and right ends of the chain. The system is therefore kept out-of-equilibrium by this imbalance. If we set the two temperatures to the same value, then a thermal equilibrium is established in the long time regime \cite{Mejia-Monasterio2007}. 

\begin{figure} 	
	\includegraphics[width=10cm]{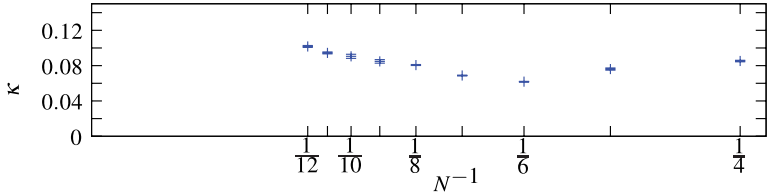}
	\caption{Thermal conductivity of the integrable spin chain as a function of the inverse of the chain length, $N$. For large $N$ it seems the conductivity starts to grow suggesting the establishment of a ballistic thermal transport regime.  System parameters: $\beta_L=0.25$, $\beta_R = 0.5$, $Q=\lambda= 0.01$, $\omega= 1$. Simulation parameters: $\Delta t = 1$, $t_{fin} = 10^7$ for $N\geq 9$ or $t_{fin} = 10^8$ for $N\leq 8$. From \cite{Mejia-Monasterio2007}, reprinted with permission.}
	\label{spin_therm_cond}
	\end{figure}

In figure, \ref{spin_therm_cond} the behaviour of the thermal conductivity of the spin chain as a function of the length of the chain is reported. Interestingly, the thermal conductivity starts to grow for larger system sizes, suggesting that a ballistic thermal transport regime might be reached for this chain. 


\subsection{Thermal Relaxation dynamics}\label{relaxation}

In the derivation of the SSE in section \ref{sse} we assumed the
external environment is in thermal equilibrium with temperature $T=(k_B \beta)^{-1}$,
\begin{equation}
	\hat \rho^{\mathrm{eq}}_B = \frac{e^{-\beta \hat H_B}}{\Tr_B e^{-\beta \hat H_B }}.
\end{equation}
From basic thermodynamical considerations \cite{Feynman,Huang2001},
we expect the open quantum system, which is in contact with this equilibrated
heat bath, to evolve in the long-time to some
steady state that coincides with its thermal equilibrium at the same temperature
$T$ as the heat bath,
\begin{equation}
	\lim_{t\rightarrow \infty}\hat \rho_S(t) = \frac{e^{-\beta \hat H_S }}{\Tr_S e^{-\beta \hat H_S }}.
\end{equation}
This dynamics towards thermal equilibrium is a non-equilibrium process. 
In order to describe realistic open systems and hence
introduce the temperature in the description of the dynamics,
the equation of motion should ensure, in the long time, this relaxation behaviour. If one considers
the approximations performed in the derivation of the SSE, this begs the question
of whether the non-Markovian SSE is still able to describe thermal relaxation processes
on average.

In order to understand and investigate these processes, one is interested
in the conditions for thermal relaxation and how they enter the NMSSE (\ref{eq:NMSSE}),
\begin{eqnarray}
  	i d |\phi(t)\rangle&=& \hat H_S |\phi(t)\rangle dt +\lambda  \gamma(t)  \hat V |\phi(t)\rangle dt\nonumber\\
	&& - i \lambda^2 \hat V \int_0 ^t d \tau  e^{- i \hat H_S\tau} \hat V  C(\tau)|\phi(t-\tau)\rangle dt.
\end{eqnarray}
Here, we consider the environment as described by one operator as this case contains all the
essential characteristics we would like to discuss. The generalization to an environment
described by many operators requires some additional care. If the same temperature is uniform,
we do expect the system to reach a thermal equilibrium at the given temperature. If
temperature is not uniform, we should not expect any thermal relaxation: the system is
continuously driven out-of-equilibrium by the temperature gradients between the different
parts of the environment. One can see that the coupling of the system with the environment
enters the equation of motion only through two quantities, namely the bath correlation
function $C(\tau)$, which is connected with the noise, and the system's coupling operator
$\hat V$. Whether the system is driven towards thermal equilibrium by the coupling to the heat
bath can thus only depend on these two quantities. However, one naturally expects that the
operator $\hat V$ from the reduced Hilbert space of the system does not contain information
about the environment, like for example its temperature. Hence, thermal relaxation processes
have to be highly dependent on the structure of the bath correlation function $C(\tau)$ as it
is the only quantity that describes the coupling from the side of the environment.

As we have seen before, the SSE describes an ensemble of wave functions evolving under the
influence of different stochastic processes. Consequently, only on average one will be able to
state whether thermal equilibrium is reached and thus we will use the in section
\ref{density_matrix} derived master equation (\ref{eq:master_2n_order}) for the discussion of
thermal relaxation processes. First of all, we are interested in whether or not thermal
equilibrium is reached and thus it is sufficient to investigate this with the help of the
Redfield master equation (\ref{eq:REDF}) in the limit $t\rightarrow \infty$. One expects,
indeed, the master equation (\ref{eq:master_2n_order}), which corresponds to the NMSSE, and
Redfield equation (\ref{eq:REDF}) to have the same steady states. In order that the thermal
equilibrium state of the system is a stationary state of the non-Markovian SSE, the following
equation has to be satisfied
\begin{equation}\label{eq:cond_th_eq}
	\lim_{t\rightarrow \infty}\frac{d \hat{\rho}^{\mathrm{eq}}_S(t) } {d t} = 0.
\end{equation}
This requirement and the fact that the equilibrium density operator commutes
with the system Hamiltonian lead to a condition for the NMSSE to ensure thermal relaxation,
\begin{equation}
\label{eq:condition_1}
0=\lim_{t\rightarrow \infty}\frac{d \hat{\rho}^{\mathrm{eq}}_S}{dt}=\hat{K}\hat{\rho}^{\mathrm{eq}}_S \hat V + \hat V  \hat{\rho}^{\mathrm{eq}}_S \hat{K}^{\dagger}-\hat V\hat{K}\hat{\rho}^{\mathrm{eq}}_S -
\hat{\rho}^{\mathrm{eq}}_S \hat{K}^{\dagger}\hat V+\mathcal{O}(\lambda^4),
 \end{equation}
where
\begin{equation}\hat{K}=\lambda^2\int_0^{\infty} d  \tau C(\tau) e^{- i \hat H_S\tau} \hat V e^{i \hat H_S\tau} .
\end{equation}
From this one can obtain the conditions for stochastic relaxation processes,
to this end, we will change to the energy basis of the system, 
\begin{equation} \label{eq:energy_basis}
	\hat H_S |n \rangle =\epsilon_n  |n \rangle ,
	\hat V=\sum_{n,m} v_{nm} |n \rangle \langle m | ,
\end{equation}
in which (\ref{eq:condition_1}) can be written as
\begin{eqnarray}
\label{eq:condition_2}
\lim_{t\rightarrow \infty}\frac{d \hat{\rho}^{\mathrm{eq}}_S}{dt}=& \lambda^2 \sum_{n,m,l} |n\rangle \langle l | v_{nm} v_{ml} \\ 
&\times \int_0^{\infty} d  \tau\bigg\{C(\tau)e^{-i(\epsilon_n-\epsilon_m)\tau}e^{-\beta \epsilon_m} + C^{\ast} (\tau)e^{-i(\epsilon_m-\epsilon_l)\tau}e^{-\beta \epsilon_m}\nonumber \\
& -C(\tau)e^{-i(\epsilon_m-\epsilon_l)\tau}e^{-\beta \epsilon_l}- C^{\ast} (\tau)e^{-i(\epsilon_n-\epsilon_m)\tau}e^{-\beta \epsilon_n} \bigg\}+\mathcal{O}(\lambda^4)\nonumber.
\end{eqnarray}
As discussed before we are interested in system independent conditions under
which the right hand side of (\ref{eq:condition_2}) vanishes. Hence,
these have to be connected with the bath correlation function,
\begin{equation}
C(\tau)=\Tr_B\big[\hat \rho^\mathrm{eq}_B \hat B(\tau)\hat B(0)\big],
\end{equation}
and by using the fact that $\hat B$ is a hermitian operator, one can conclude that
\begin{equation}\label{eq:prob_C}
	C^{\ast}(\tau)=C(-\tau) .
\end{equation}
As a result of this, the \lq half Fourier transform\rq\ in (\ref{eq:condition_2})
can be written as
\begin{eqnarray} \label{eq:int_rewritten}
\int_0^{\infty} d  \tau C(\tau) e^{-i \omega \tau}&=& \frac{1}{2}  \int_{-\infty}^{\infty} d  \tau C(\tau) e^{-i\omega
\tau}\nonumber\\
&& + i\frac{1}{2i} \int_0^{\infty} d  \tau \bigg\{ C(\tau)e^{-i \omega \tau} -C^{\ast}(\tau)e^{i \omega \tau} \bigg\}\nonumber \\
&=&\frac{1}{2}\widehat{C}(\omega) + iD(\omega) , 
\end{eqnarray}
where $D(\omega)$ is the imaginary part of this half Fourier transform and as a consequence
the Fourier transform of the bath correlation function, $\widehat{C}(\omega)$,
is a real-valued function. Furthermore, if the half Fourier transform (\ref{eq:int_rewritten})
is analytic in the upper complex half plane of $\omega$ and vanishes faster than
${|\omega|^{-1}} $ as $\omega$ goes to infinity, one can apply the Kramers-Kronig relation,
\begin{equation}
D(\omega)=\frac{1}{2\pi}P \int_{-\infty}^{\infty} d  a \frac{ \widehat{C}(a) }{\omega-a}.
\end{equation}
Here, $P \int$ denotes the Cauchy principal value of the integral. 

In the same spirit the half Fourier transform of the conjugate bath correlation function can be written as
\begin{eqnarray} \label{eq:int_rewritten_both}
 \int_0^{\infty} d  \tau C^{\ast}(\tau) e^{-i \omega \tau} &=& \widehat{C}(-\omega) + i\frac{1}{2\pi}P\int_{-\infty}^{\infty} d  a\frac{ \widehat{C}(-a) }{w-a} \nonumber\\
&=& \frac{1}{2}\widehat{C}(-\omega)+ i F(\omega) .
\end{eqnarray}
Inserting equations (\ref{eq:int_rewritten_both}) and (\ref{eq:int_rewritten}) into
(\ref{eq:condition_2}) leads to 
\begin{eqnarray}
\label{eq:condition_333}
\lim_{t\rightarrow \infty}	\frac{d \hat{\rho}^{\mathrm{eq}}_S}{dt}&=& \lambda^2 \sum_{n,m,l} |n\rangle \langle l |v_{nm} v_{ml} \Bigg\{  \Big( \frac{1}{2} \widehat{C}(\omega_{nm})+ i
 	D(\omega_{nm})\Big) e^{-\beta \epsilon_m}  \nonumber\\
  	&&+\Big(\frac{1}{2} \widehat{C}(-\omega_{ml})+ i F(\omega_{ml})\Big)e^{-\beta\epsilon_m}  \nonumber\\
    &&- \Big(\frac{1}{2} \widehat{C}(\omega_{ml})+ i D(\omega_{ml})\Big)e^{-\beta\epsilon_l}  \nonumber\\
    &&-\Big(\frac{1}{2} \widehat{C}(-\omega_{nm})+ i F(\omega_{nm})\Big)e^{-\beta\epsilon_n}  \Bigg\}+\mathcal{O}(\lambda^4) , 
\end{eqnarray}
where $\omega_{ij}= \epsilon_i-\epsilon_j$. We want to point out that in order to satisfy
the requirement of thermal relaxation, the right-hand side of this equation has to vanish.
In addition, the Fourier transform of the bath correlation function can be interpreted as
the power spectrum of the noise and hence describes the probabilities for energy transitions
in the system. By assuming that this power spectrum satisfies a so-called detailed-balance relation,
\begin{equation}\label{eq:detailed_balance}
  \widehat{C}(-\omega)= e^{\beta \omega} \widehat{C}(\omega) ,
\end{equation}
(\ref{eq:condition_333}) simplifies to 
\begin{eqnarray} \label{eq:condition_4}                                   
 \lim_{t\rightarrow \infty}	\frac{d \hat{\rho}^{\mathrm{eq}}_S}{dt}&=& i\lambda^2 \sum_{n,m,l} |n\rangle \langle l | s_{nm} s_{ml} \Bigg\{  D(\omega_{nm}) e^{-\beta \epsilon_m} +
	F(\omega_{ml})e^{-\beta\epsilon_m} \nonumber\\
	&&-  D(\omega_{ml})e^{-\beta\epsilon_l} -  F(\omega_{nm})e^{-\beta\epsilon_n} \Bigg\}+\mathcal{O}(\lambda^4) .
\end{eqnarray}
The detailed-balance relation (\ref{eq:detailed_balance}) ensures that the energy transitions
in the system are balanced according to Boltzmann statistics. Furthermore, we want to point out
that (\ref{eq:condition_4}) has only imaginary components and by inserting the explicit
integrals into (\ref{eq:condition_4}) we arrive at
\begin{eqnarray}   \label{eq:condition_5}                                       
 	\lim_{t\to\infty}\frac{d \hat{\rho}^{\mathrm{eq}}_S}{dt}&=& \frac{i\lambda^2}{2 \pi} \sum_{n,m,l} |n\rangle \langle l | v_{nm} v_{ml} \int_{-\infty}^{\infty} d  a
\Bigg\{ \frac{\widehat{C}(a)\big(1-e^{-\beta(\epsilon_n-\epsilon_m-a)} \big)e^{-\beta \epsilon_m}}{\omega_{mn}-a}\nonumber\\
&&- \frac{\widehat{C}(a)\big(1-e^{-\beta(\epsilon_m-\epsilon_l-a)} \big)e^{-\beta \epsilon_l}}{\omega_{ml}-a}\Bigg\} + \mathcal{O}(\lambda^4) .
\end{eqnarray}
In this expression there is no need to write the principal value anymore since the integral
is no longer singular. It can be shown that the diagonal components of (\ref{eq:condition_5})
cancel each other, i.e.,
\begin{equation}
 	\langle l| \frac{d \hat{\rho}^{\mathrm{eq}}_S}{dt} |l \rangle = 0.
\end{equation}
This can be done by changing $\omega_{mn}$ to $-\omega_{mn}$, changing the integration variable
in the second integral of (\ref{eq:condition_5}) from $a$ to $-a$, and applying the detailed-balance
relation (\ref{eq:detailed_balance}) again.

As a result, one can conclude that the NMSSE  (\ref{eq:NMSSE}) has a stationary solution which coincides with the
thermal-equilibrium state up to first order in $\lambda$. Furthermore, if the detailed-balance relation
is satisfied, the corresponding master equation (\ref{eq:REDF}) drives the system towards a stationary
state that coincides in the diagonal elements in the energy basis with the thermal-equilibrium state up
to fourth order. Additionally, neglecting either the off-diagonal components of the density matrix in
the long-time behaviour or the imaginary contribution of the half Fourier transform,
\begin{equation}
	\mathrm{Im} \left[ \int_0^{\infty} d  \tau C(\tau)e^{-i \omega \tau}\right] = D(\omega) \approx 0,
\end{equation}
the system will be driven towards equilibrium if the equation is considered up to fourth order.
It can be argued that this imaginary part can be included in the
system Hamiltonian \cite{Weiss2007}, leading to the so-called Lamb shift. This means that the
equilibrium density operator will not commute with this effective Hamiltonian. Nevertheless,
this energy shift will not introduce dissipative dynamics in the system \cite{Breuer2002}.

In conclusion, the bath correlation function should satisfy
the detailed-balance relation in order to describe thermal-relaxation behaviour.
Hence, by the Markovian approximation, $C_{\alpha,\beta}(\tau)\approx D_{\alpha,\beta}\delta(\tau)$,
one neglects this property of the bath correlation function. Therefore, the description of thermal
relaxation processes with the help of the Markovian SSE seems to be questionable.

As an example to demonstrate the previous considerations about thermal relaxation processes
we will discuss a simple model. Namely, a tight-binding model with three sites,
\begin{equation}	\label{tight-binding}
	\hat H_S = -\Gamma (\hat c^{\dagger}_1 \hat c_2 + \hat c^{\dagger}_2 \hat c_1 + \hat c^{\dagger}_2 \hat c_3 + \hat c^{\dagger}_3 \hat c_2 ),
\end{equation}
where the operators $\hat c^{\dagger}_i$ create an electron at the $i$-th tight-binding position
and $\Gamma$ is the hopping integral. We will couple this electronic system to the electromagnetic
field inside a cavity. For this model one can derive the power spectrum in dipole approximation ($c=1$),
\begin{equation}
	\widehat{C}_{cavity}(\omega)=\frac{|\omega|^3}{\pi \Omega \epsilon_0} \left[f(\beta,|\omega|)+\theta(-\omega)\right]~\mbox{if}~|\omega|<\omega_{c},
	\label{cavity-power-spectrum}
\end{equation}
where $f(\beta,\omega)=1/(\exp(\beta \omega)-1)$, $\theta(\omega)$ the Heaviside function, $\omega_c$ is a cutoff
frequency (determined by the dimensions of the system to satisfy the dipole
approximation) and $\Omega$ is the volume of the cavity. For $|\omega|>\omega_c$ this
function vanishes. We want to point out that this power spectrum contains naturally
the detailed-balance relation and hence the model is suitable to describe thermal
relaxation processes.

In addition, one can derive the system's coupling operator for the considered model as
\begin{equation}
	\hat V = -q \sum_{l,p}\vec{u} \cdot \langle l | \vec{r} | p \rangle  \hat \epsilon_l^{\dagger} \hat \epsilon_p ,
\end{equation}
where $q$ is the electron charge and the operator $\epsilon_l^{\dagger}$
creates an electron in the $l$-th energy eigenstate of the system.
Here, we have assumed for simplicity that each mode of the cavity
has the same polarisation direction $\vec{u}$, parallel to the chain
of the three tight-binding sites. We want to point out that the form
of this operator should be immaterial for the establishment of a thermal
equilibrium, since this is only determined by the power spectrum (\ref{cavity-power-spectrum}).
Moreover, if the same operator is used in a Markov approximation,
a steady state is established that is not the equilibrium configuration.

\begin{figure}[htbp]
	\centering
		\includegraphics[width=8cm]{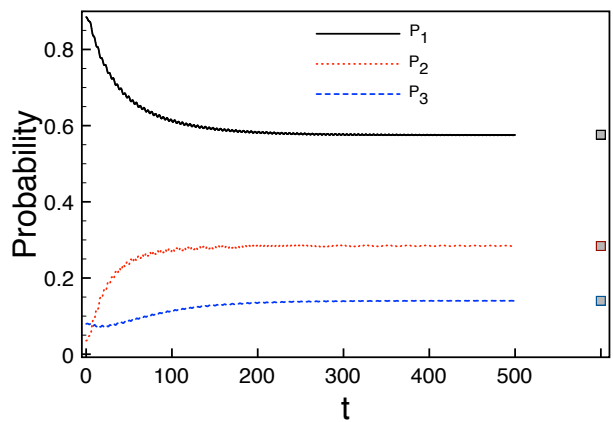}
	\caption{
The probability of occupying the three eigenstates of the Hamiltonian (\ref{tight-binding}) as
a function of time obtained from the numerical solution of (\ref{eq:master_2n_order}). The
system is driven towards thermal equilibrium, represented by the three final points in the
figure, in the long time regime.}
	\label{probability3x3}
\end{figure}

In figure~\ref{probability3x3} we report the probability of occupying the
three eigenstates of the Hamiltonian (\ref{tight-binding}) as a function of time obtained from the NMME (\ref{eq:master_2n_order}). 
For this dynamics, we have used the parameters, $\beta=1$,
$\omega_c=1$, $\Gamma=1$ and $\lambda=0.1$. We have used a simple Euler algorithm
\cite{Press1992} with a time step $\delta t=0.001$ to numerically
solve the NMME (\ref{eq:master_2n_order}). 
Unfortunately, the solution of the NMSSE and master equation requires the evaluation of the
history integral which contains the correlation function. This integral needs to be evaluated
at each time step, therefore making the solution of the NMSSE unattainable in a reasonable
amount of time. This is because one needs to build some statistics before performing the
averages: for obtaining a reasonable colored noise following the discussion in
\ref{solving_the_sse} we need to average over thousands of runs. As we discuss elsewhere
\cite{Biele2011}, one can approximate the NMSSE with an equivalent equation up to the fourth
order in $\lambda$. This approximation reduces the numerical needs to solve the NMSSE, since
we arrive at a NMSSE that is local in time. We are able to show that, this new SSE (still
non-Markovian) and the corresponding master equation give the same dynamics. This bring the
numerical cost of solving the NMSSE to the same level with the Markovian SSE \cite{Biele2011}.

\subsection{Ionic motion}
\begin{figure}[ht!]
	\includegraphics[width=12cm]{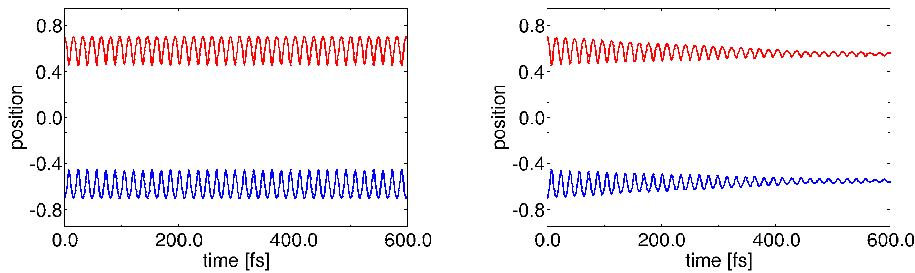}
	\caption{The (classical) dynamics of the ionic positions of a Neon dimer. On the left, the system is closed and the dynamics periodic, on the right, the ions dissipate energy to the electrons in contact with an external environment. The system therefore reaches the ground state after some time. From \cite{Appel2011}, reprinted with permission.}  
	\label{heiko}
\end{figure}

An interesting extension of the result of the preceding section \ref{stochastic_dft} on DFT
has been proposed recently by H. Appel and M. Di Ventra \cite{Appel2009}. They used the
electrons present in a certain molecule as the ``link'' between the ions and the external
environment, therefore investigating how some vibrational modes, excited at the initial time
and coupled to the electron motion can be effectively dissipated \cite{Appel2009}. The
starting point is a theory similar to the one we have discussed in section
\ref{stochastic_dft} where one considers the total current and charge densities, namely,
including in those expressions also the ionic contribution. In a system of electrons and ions,
where the former are connected to external baths, one proves that the total electrical current
density is in a one-to-one correspondence to the external vector potential, thus establishing
a stochastic TD-DFT scheme for the system under investigation. This clearly allows for energy
dissipation from the ionic degrees of freedom via the electrons. In figure~\ref{heiko}, the
dynamics of the ionic positions of a Neon molecule is shown \cite{Appel2011}. In the left
pane, the electrons are disconnected from the environment, and the system undergoes a periodic
motion determined by the initial conditions. In the right pane, the electrons are connected
with an external bath that continuously extracts energy from them: the ions relax to the
equilibrium configuration of the Neon dimer, with the electrons in the ground state.
It is clear that this is a simple example of what is an interesting and promising technique to investigate the correlated dynamics of electrons and ions coupled to an external environment. For example, this dissipative dynamics could possibly explain the appearance of broadening on the electron and phonon spectrum.



\subsection{Bose-Einstein Condensation}\label{BEC}                                    

As a last example of application of the stochastic Schr\"odinger equation we consider again
the Bose-Einstein condensation \cite{Huang2001,Feynman1963} of Alkali gases
\cite{Dalfovo1999}. In this case, we depart from the SSE we have discussed in previous
sections, and introduce the case of particle exchange between the system and its environment.
This allows us to investigate the dynamics of the condensate formation and predict the
condensation temperature. To do so, we will need to introduce a stochastic term that is not
``multiplicative'', i.e., it does not multiply the wavefunction as happens in \ref{sse}, but
``additive'' as we will discuss in a moment.

In the experimental realization of the condensate, the gas is trapped into a confining
harmonic potential generated by a magnetic field obtained by focusing two laser fields
\cite{Davis1995, Anderson1995}. In these systems the final temperature is about a few hundreds
nano-Kelvin. To reach this fantastic result, a novel technique was developed, the evaporative
cooling of the gas. At temperatures above the condensation temperature, the gas follows a
Boltzmann statistics. The evaporative cooling selectively removes from the gas the particles
with the highest kinetic energy. This is possible since the highest kinetic energy states are
those who have a finite probability density at the border of the magnetic trap. Therefore,
another laser focused around the boundaries of the magneto-optical trap can allow for those
bosons to be released from the trap. The gas, where the tail of the Boltzmann distribution has
been cut, relaxes via boson-boson collisions to a new distribution with smaller temperature,
since the total kinetic energy is lower than before. By selectively removing the hottest
atoms, the temperature of the Alkali gas can be reduced below the condensation temperature and
one can observe a condensation both in momentum and position space
\cite{Dalfovo1999,Davis1995,Anderson1995}.

As we have rapidly discussed in section \ref{equivalence}, the Gross-Pitaevskii (GP) equation
is the standard theoretical tool to describe the physics of the condensate system
\cite{Dalfovo1999}. The equation has proven successful in describing both for the dynamics and
static properties of the condensate. Due to its simplicity and striking successes, many people
have tried to generalize the GP equation to study the dynamics of the condensate formation. In
this system, the ground state and the excited states are well separated in energy --this
separation is dictated by the confining potential-- and one could regard the excited states as
a reservoir of particles and energy for the condensate. With an expansion on the
particle-particle interaction, by using the Keldish formalism Stoof \cite{Stoof1999} arrived
at the following stochastic GP equation
\begin{eqnarray}\label{stochastic-gpe}
	i\partial_t \Phi(r,t)&=&\left[-\frac{\nabla^2}{2m}+V_{ext}(r)-\mu-iR(r,t)+g|\Phi(r,t)|^2\right]\Phi(r,t)\nonumber\\
	&&+\eta(r,t).
\end{eqnarray}     
Here, $\Phi(r,t)$ describes the condensate phase,
$V_{ext}$ is the confining potential, $\mu$ is the chemical potential, $g$ the interaction
constant. Finally, $R(r,t)$ and $\eta(r,t)$ are the terms coming from the coupling of the
condensate to the non-condensate atoms which form the ``environment'' for the condensate.
Notice that $\eta(r,t)$ does not multiply the wavefunction $\Phi(r,t)$, but is added to the
equation of motion. For this reason, $\eta$ is an additive noise for our problem. This is
necessary in order for (\ref{stochastic-gpe}) to allow particle exchange with the environment,
i.e, the norm of the function $\Phi(r,t)$ can vary with time. Clearly, the term $R(r,t)$
represents a dissipative contribution denoting the coupling between the condensate and the
non-condensate phase, while $\eta(r,t)$ accounts for the incoherent collisions within the gas,
analogously to the description of the Brownian motion. Interestingly, one can arrive at a
similar equation with a completely different technique. This technique, developed in parallel
and independently from the one we present here, looks promising in building a theory to
investigate the dynamical formation of the Bose-Einstein condensate in Alkali gases
\cite{Gardiner2003,Wright2008,Blakie2008}. Its first application to the investigation of
vortex formation close to the transition temperature has provided results in good quantitative
agreement with the experiments \cite{Weiler2008}. Here, however, we do not feel necessary to
expose this theory. The interested reader could find a clear exposition in
\cite{Gardiner2003}.

The noise $\eta$ is characterized by a Gaussian correlation function
\begin{equation}
	\overline{\eta^*(r',t')\eta(r,t)}=\frac{i}{2}\Sigma^k(r,t)\delta(t-t')\delta(r-r'),
\end{equation}
where $\Sigma^k(r,t)$ is the Keldish self-energy of the gas of Alkali atoms \cite{Stoof1999,Stoof2001,Cockburn2009}. As one could
expect, there is an equivalent form of the dissipation-relaxation relation that connects the dissipation term $R$ to the noise
$\eta$:
\begin{equation}
	iR(r,\epsilon)=-\frac{\Sigma^k(r,\epsilon)}{2(1+2f(\beta,\epsilon))},  
	\label{bec_dissipative}
\end{equation} 
where $R(r,\epsilon)$ and $\Sigma^k(r,\epsilon)$ are the Fourier transform of $R(r,t)$ and $\Sigma^k(r,t)$, respectively and $f(\beta,\epsilon)$ is the Bose-Einstein distribution at temperature $T$.    
                                   
\begin{figure}[htbp]
	\centering
		\includegraphics[width=10cm]{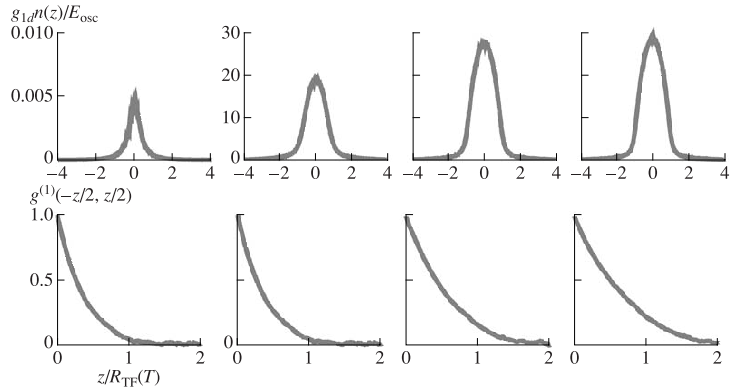}
		\caption{(Upper panel) Snapshots of the particle density during the growth process of a one-dimensional quasi-condensate of $^{23}$Na.
For the simulations, the parameters $\mu=30 \times E_{osc}$ where $E_{osc}=\hbar \omega_z$ with $\omega_z= 2\pi\times
30$ Hz for a final quasi-condensate of $N=20000$ particles and a final temperature of $T=100$ nK. Displayed snapshots are
at $t/t_{eq} = 0.05$, $0.15$, $0.30$, and $0.45$, where $t_{eq}$ is an approximate time for the system to reach dynamical
equilibrium. The effective interaction $g_{1d}$ is obtained after averaging over a transverse Gaussian profile of width
$l=\sqrt{\hbar/m\omega_T}$ where $\omega_T=2\pi\times 120$ Hz. (Lower panel) The correlation function for the quasi-condensate. When the quasi-condensate is forming, leftmost plot, the correlation function is expected to decay exponentially with the distance. When the quasi-condensate is formed, rightmost plot, the correlation function decays as a power law. Reprinted from \cite{Cockburn2009} with permission.}
	\label{cockburn}
\end{figure}

The dynamics of the condensate formation can be obtained numerically in one dimension. In 2D and 3D, while in principle the equations are still valid, their numerical integration is more difficult. Besides the obvious numerical overhead associated with dimensionality, this is due to the more complex spectrum of the system. Although strictly speaking in one dimension the Bose-Einstein condensation for an ideal gas does not exist \cite{Huang2001,Feynman}, in practice one considers the formation of the so-called quasi-condensate which corresponds to a macroscopic occupation of the ground state of the confining potential. To obtain an effective equation of motion, one splits the wavefunction into a radial and transverse component. Then the radial component is assumed to be Gaussian and integrated out. The process gives rise to an effective one-dimensional interaction constant, $g_{1d}$. This approximation works in an excellent way for the cigar-shaped potentials \cite{Dalfovo1999} where the confinement along the radial component is stronger than along the transverse direction. In figure \ref{cockburn} we report the dynamics of formation of the quasi-condensate at different times. At the end of the process, a quasi-condensate of 20000 particles is formed at a temperature of about 100 nK. 
A key quantity to establish the presence of a phase transition is the so-called first-order correlation function, defined as
\begin{equation}
	g^{(1)}(-z/2,z/2;t)=\frac{\overline{\Phi^*(-z/2,t)\Phi(z/2,t)}}{\overline{|\Phi^*(-z/2,t)|^2}~\overline{|\Phi(z/2,t)|^2}}.
\end{equation}
This quantity can be measured experimentally \cite{Richard2003,Hellweg2003,Cacciapuoti2003}, thus providing a test for the theory. 
In figure \ref{cockburn}, distances are scaled to the quasi-condensate dimension at equilibrium.
The effective radius $R_{TF}(T)$ varies with temperature starting from the Thomas-Fermi expression at zero temperature
$R_{TF}(0)=\sqrt{2\mu/m\omega_z^2}$ if $\omega_z$ is the frequency of the confining harmonic potential in the $z$
direction, and $\mu$ is the chemical potential, i.e., the energy of the condensate. For the discussion on how to
calculate $R_{TF}$ at finite temperature see (\cite{Cockburn2009}) and references therein.

The results of figure \ref{cockburn} can be compared with the experimental results obtained
since 1995, the year of the first experimental observations of the Bose-Einstein condensate in
ultra-cold Alkali atomic gases \cite{Davis1995,Anderson1995}. The theoretical modelling is
able to describe the quasi-condensate formation with a throughout investigation of the
dynamics. It is also important that the theory contains essentially no fitting parameters,
since all of them can be quantitatively determined from the knowledge of the experimental
setup.

The Bose-Einstein condensation in Alkali gases has proven to be an important tool for a
deeper understanding of quantum mechanical phenomena. In particular, the relative easiness in
manipulating it and its relative stability have allowed for a series of interesting
experiments on quantum measurement theory to be carried out. Indeed, the condensate can allow
for either invasive or non-destructive probe of certain physical quantities. In this respect,
it has been recently possible to develop a stochastic theory to describe the case in which the
condensate is interacting with a laser field which is under continuous measurement. As is the
case with the theory of continuous quantum measurement in other systems, this theory do not
fall into the form of the SSE we have extensively presented in this review. The interested reader could start from \cite{Dalvit2002,Szigeti2010}.


\section{Summary}\label{conclusion}
The stochastic Schr\"odinger equation is a ubiquitous tool that can be used to describe the
dynamics of a open quantum system, namely a system in interaction with an external
environment. The accurate use of this tool requires a deep understanding of the physical
approximations used to derive, from the Schr\"odinger equation of the total system, an
effective equation of motion for the subsystem of interest. In this review we have mainly
focused only on one class of SSE, those that have a quite specific form and structure, derived
from a few assumptions on the dynamics of the environment and on the coupling between the
system and its environment. To give to the reader a feeling of how wide is the range of
applicability of these concepts, we have briefly discussed in section \ref{BEC} a SSE with a
different form, in particular with an additive noise in contrast to the multiplicative noise
we have discussed in the rest of this review. Other stochastic equations have been suggested
in the past \cite{Razavy2006} to describe an effective dynamics of the system, without a clear
step-by-step derivation of the equation of motion. We have not considered that class here. The
reason for that is simple: each of these models requires a proper introduction, justification,
and derivation, while we are more interested in presenting a unified, although restricted,
point of view. The interested reader can start from \cite{Razavy2006} and explore the vast
literature present there. To make connection with the knowledge of the reader, we have clearly
established the link between the SSE and a master equation for the reduced density matrix.
From this point of view, one can see the SSE as a ``Langevin'' type of equation and the master
equation as its ``Fokker-Planck'' counterpart. We have shown however, that the two theories
can lead to significantly different results for physical quantities.

We have applied the SSE to different systems, from spin thermal transport, to Bose-Einstein
condensation to thermal equilibrium in electronic systems. The list of examples could go
further but we will find ourselves in repeating the same, useful, concepts.

Finally, we have discussed an interesting new topic in SSE, i.e., the inclusion of
particle-particle interactions and its application to describe the dynamics of electronic
devices. We have shown, in parallel with the DFT theory, how one can build an effective
description of the open system that in the future should allow us to construct an {\it
ab-initio} modelling of electronic thermal transport, without relying too much on any
assumption about the properties of the steady-state.

The field is still in its infancy. For example, a clear connection between the non-Markovian
SSE and its master equation counterpart is missing: the actual derivation by Gaspard and
Nagaoka \cite{Gaspard1999} is incomplete and leaves many questions open, first of all the
non-uniqueness of the derived equation. A deeper understanding of the condition for the
establishment of a steady-state or ``thermal equilibrium'' is also missing. The ``detailed
balance'' is indeed not enough to ensure the system is driven towards a ``thermal
equilibrium'', at least in the form known from standard statistical mechanics. Also, a
thorough exploration of the effect of particle-particle interaction, even at the basic level
of mean-field theories, is lacking. We therefore foresee an interesting future for these
techniques.

\ack
R. B. and R. D'A. acknowledge support from MICINN (FIS2010-21282-C02-01 and PIB2010US-00652), the Grupos Consolidados UPV/EHU del Gobierno Vasco (IT-319-07) and ACI-Promociona (ACI2009-1036). R. B. acknowledges financial
support from IKERBASQUE, Basque Foundation for Science. R. D'A. acknowledges the financial support of the MICINN CONSOLIDER-INGENIO 2010 ``NanoTherm" (CSD2010-00044) and is thankful for its hospitality to the Physics
Department of the King's College London.

\newpage
\section*{References}
\bibliographystyle{phjcp}
\bibliography{library}

\end{document}